\documentclass{aa} 
\usepackage{float}
\usepackage{hyperref}
\usepackage[T1]{fontenc}
\usepackage{epstopdf}
\usepackage{natbib}
\usepackage{soul}

\usepackage{graphicx}
\usepackage{txfonts}
\usepackage{xcolor}

\begin{document}

   \title{Debris discs in binaries: morphology and photometric signatures}

   \author{P.Thebault
          \inst{1}
          \and
          Q.Kral\inst{1}
          \and
          J. Olofsson\inst{2,3}
          }
   \institute{LESIA-Observatoire de Paris, UPMC Univ. Paris 06, Univ. Paris-Diderot, France
    \and
    Instituto de F\'isica y Astronom\'ia, Facultad de Ciencias, Universidad de Valpara\'iso, Av. Gran Breta\~na 1111, Playa Ancha, Valpara\'iso, Chile
    \and
    N\'ucleo Milenio Formaci\'on Planetaria - NPF, Universidad de Valpara\'iso, Av. Gran Breta\~na 1111, Valpara\'iso, Chile
             }  
\offprints{P. Thebault} \mail{philippe.thebault@obspm.fr}
\date{Received ; accepted } \titlerunning{debris discs in binaries}
\authorrunning{Thebault, Kral \& Olofsson}

%

%

 
  \abstract
   {As about half of main-sequence stars reside in multiple star systems, it is important to consider the effect of binarity on the evolution of planetesimal belts in these complex systems.}
   {We aim to see whether debris belts evolving in between two stars may be impacted by the presence of the companion and whether this leaves any detectable signature that could be observed with current or future instruments.}
   {We consider a circumprimary parent body (PB) planetesimal belt that is placed just inside the stability limit between the 2 stars and use the state-of-the-art DyCoSS code to follow the coupled dynamical and collisional evolution of the dust produced by this PB belt. We explore several free parameters such as the belt's mass or the binary's mass ratio and orbital eccentricity. We use the GraTeR package to produce 2-D luminosity maps and system-integrated SEDs }
   { We confirm a preliminary result obtained by earlier DyCoSS studies, which is that the coupled effect of collisional activity, binary perturbations and stellar radiation pressure is able to place and maintain a halo of small grains in the dynamically unstable region between the 2 stars. In addition, several prominent spatial structures are identified, notably a single spiral arm stretching all the way from the PB belt to the companion star. We also identify a fainter and more compact disc around the secondary star, which is non-native and feeds off small grains from the unstable halo. Both the halo, spiral arm and secondary disc should be detectable on resolved images by instruments with capacities similar to SPHERE. The system as a whole is depleted in small grains when compared to a companion-free case. This depletion leaves an imprint on the system's integrated SED, which appears colder than for the same parent body belt around a single star. This new finding could explain why the SED-derived location $r_{disc}$ of some unresolved discs-in-binaries places their primary belt in the dynamically "forbidden" region between the 2 stars: this apparent paradox could indeed be due to overestimating $r_{disc}$ when using empirical prescriptions valid for a single star case.}
   {}

   \maketitle
%

\section{Introduction} \label{intro}

Dusty debris discs have been detected around $\sim15\%$ to $\sim30\%$ of main sequence stars in the K to A stellar-types range \citep[see reviews by][and references therein]{matthews2014,hughes2018}. The observed dust is thought to originate from the collisional grinding of much larger (and harder to detect) parent planetesimals, probably leftover from the planet formation process, either through steady-state collisional cascades or transient and more violent events \citep[e.g.][]{wyat08,theb18}. The brightest of these discs have been imaged, from scattered light observations in the visible to thermal imaging up to millimetre wavelengths, revealing that most resolved discs display pronounced spatial structures, such as rings, warps, clumps, spirals, etc. \citep{kriv10,mari20}.

As about half of main sequence stars reside in binary or higher-order multiple systems \citep{duqu91,ragh10}, the question of the influence of companion stars on the properties of debris discs is a crucial one within the more general frame of understanding planet formation in binaries \citep{theb15}.
The link between disc incidence and binarity has been investigated by \citet{tril07,rodri12,rodri15} and, more recently, in \citet{yelv19} with a sample of 341 binaries for which they look for dust-induced excesses at 70 and 100 $\mu$m. These studies have shown that, while the incidence of discs in binaries of separation $\rho \gtrsim100\,$au is similar to that of single stars, there is a strong depletion of discs in tighter binaries, with for instance no discs detected in binaries with 25\,au$\leq\,\rho\leq\,$135\,au \citep{yelv19}. This feature agrees well with a comparable deficit of younger, primordial proto-planetary discs (PPDs) in pre-main-sequence binaries of separation smaller than $\sim50-100$\,au \citep{krau12}. The fact that this critical separation is comparable to the typical radial extent of PPDs and debris discs is not surprising. Indeed, for a given binary, the whole radial domain between $\sim1/3\rho$ and $\sim3\rho$ is expected to be dynamically unstable \citep{holm99}, meaning that a $\rho\sim100\,$au binary could clear out a large fraction of the circumstellar discs around each star. However, when using dust temperature estimates to infer disc radii, \citet{tril07} found that some discs seem to be located within these dynamically "forbidden" regions. \citet{yelv19} found that 9 out of 38 disc-hosting binaries are seemingly in such an unstable configuration, but argue that these numbers should be taken with caution because of the inherent uncertainties of these disc-radii estimates.

The discovery of these puzzling unstable discs motivated \citet{theb10} and \citet[][hereafter TBO12]{theb12a} to carry out the first global studies of the dynamical response of a debris disc to a companion star. Such studies are a challenge to conventional numerical codes, because they have to take into account the binary's gravity, but also stellar radiation pressure (SRP) and the production of collisional fragments within the disc. \citet{theb10} showed that the combined effect of the latter two processes, by steadily producing small debris that can be, depending on their size, placed on unbound or highly-eccentric orbits by SRP, can feed small grains to dynamically unstable regions faster than they are ejected by binary perturbations. TBO12 expanded on this study by developing a new code, "DyCoSS", that can follow the \emph{spatial} evolution of a perturbed collisionally active disc in the presence of SRP. Preliminary results on binaries showed that the dust distribution can display spatial structures, both inside the stability region and outside of it. 

We intend to take these studies a step further by addressing several important issues that had not been dealt within these 2 pioneering studies. A first objective is to increase the spatial domain over which we follow the system's evolution, by expanding it beyond the orbit of the companion star but also to regions much closer to the primary star. We also go beyond the density maps of TBO12 and explore the disc's observed aspect at different wavelengths, ranging from the visible to millimetre. In particular, we plan on estimating if the dust that produces structures in the unstable regions can be observable on scattered light and mid-IR images, and if new features can be identified that can be unambiguously associated to binarity effects. For this, we take into account the contribution of the companion star to the dust temperature and luminosity. We also investigate if binarity leaves a detectable signature on the Spectral Energy Distribution (SED) of the system, and how our results compare to the trends identified in the global discs-in-binaries surveys, in particular that of \citet{yelv19}.

\section{Model}\label{model}

\subsection{Basic principle}

\begin{figure*}
\makebox[\textwidth]{
\includegraphics[scale=0.266]{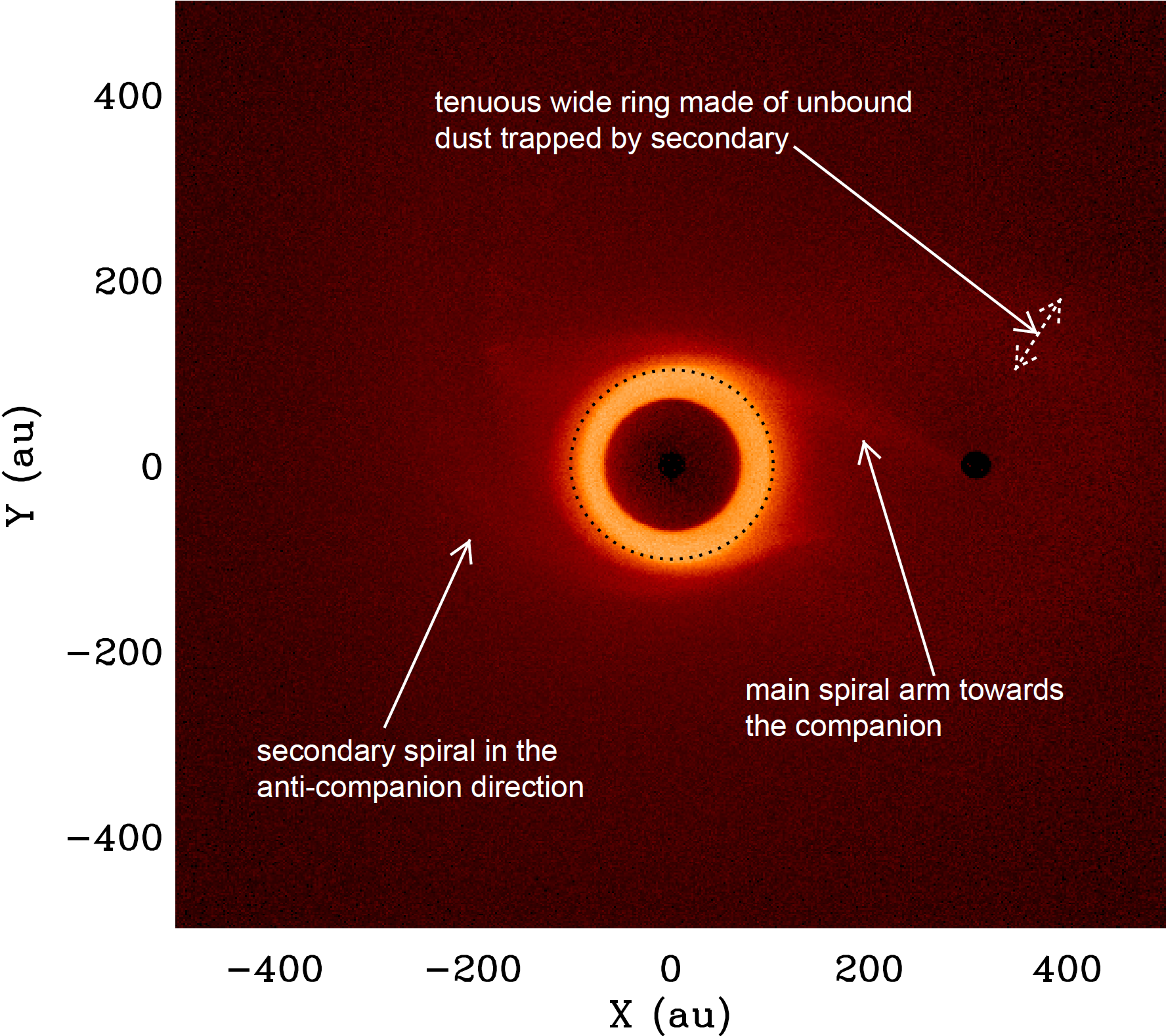}
\includegraphics[scale=0.5]{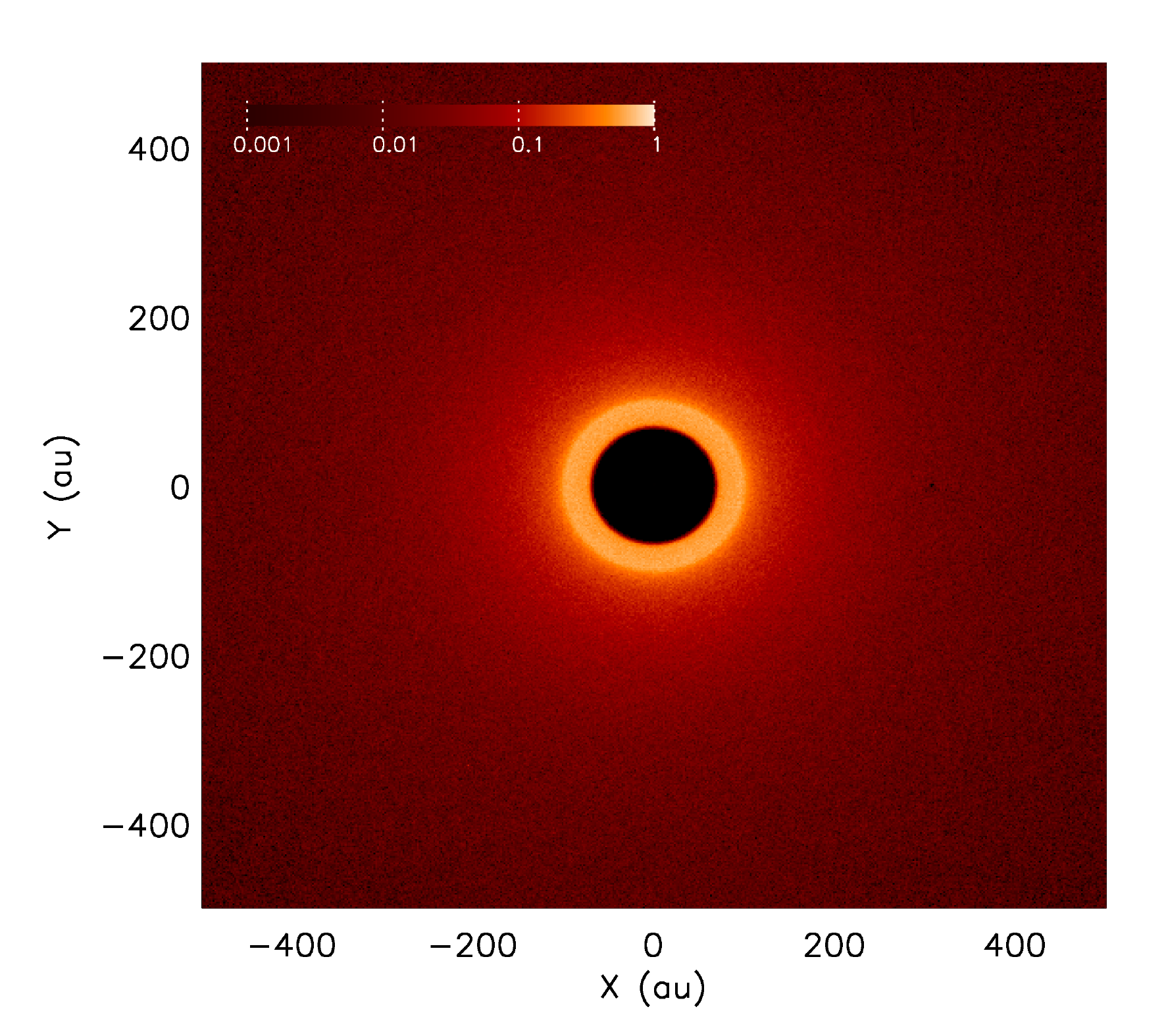}
}
\caption[]{Nominal case ($e_{\rm{b}}=0$, $\mu=0.5$ and <$\tau$>$=2\times10^{-3}$): Normalized surface density map at steady state (left). The black dotted line represents the theoretical dynamical stability limit around the primary star as derived from \cite{holm99} (An animated version of this figure, showing the disc's evolution over one binary orbit, can be found at \url{https://lesia.obspm.fr/perso/philippe-thebault/animsurf.gif}). Because accurate orbital computing requires too short timesteps when grains pass too close to the primary or the secondary, we chose to remove all particles within 14\,au of each star, hence the two black circles that mark the position of the two stars.
The right panel presents, as a comparison, the corresponding map for a disc of same initial spatial extent \emph{without} stellar companion. The relative colour scale is the same for both figures.}
\label{denscomp}
\end{figure*}

\begin{figure}
\includegraphics[scale=0.5]{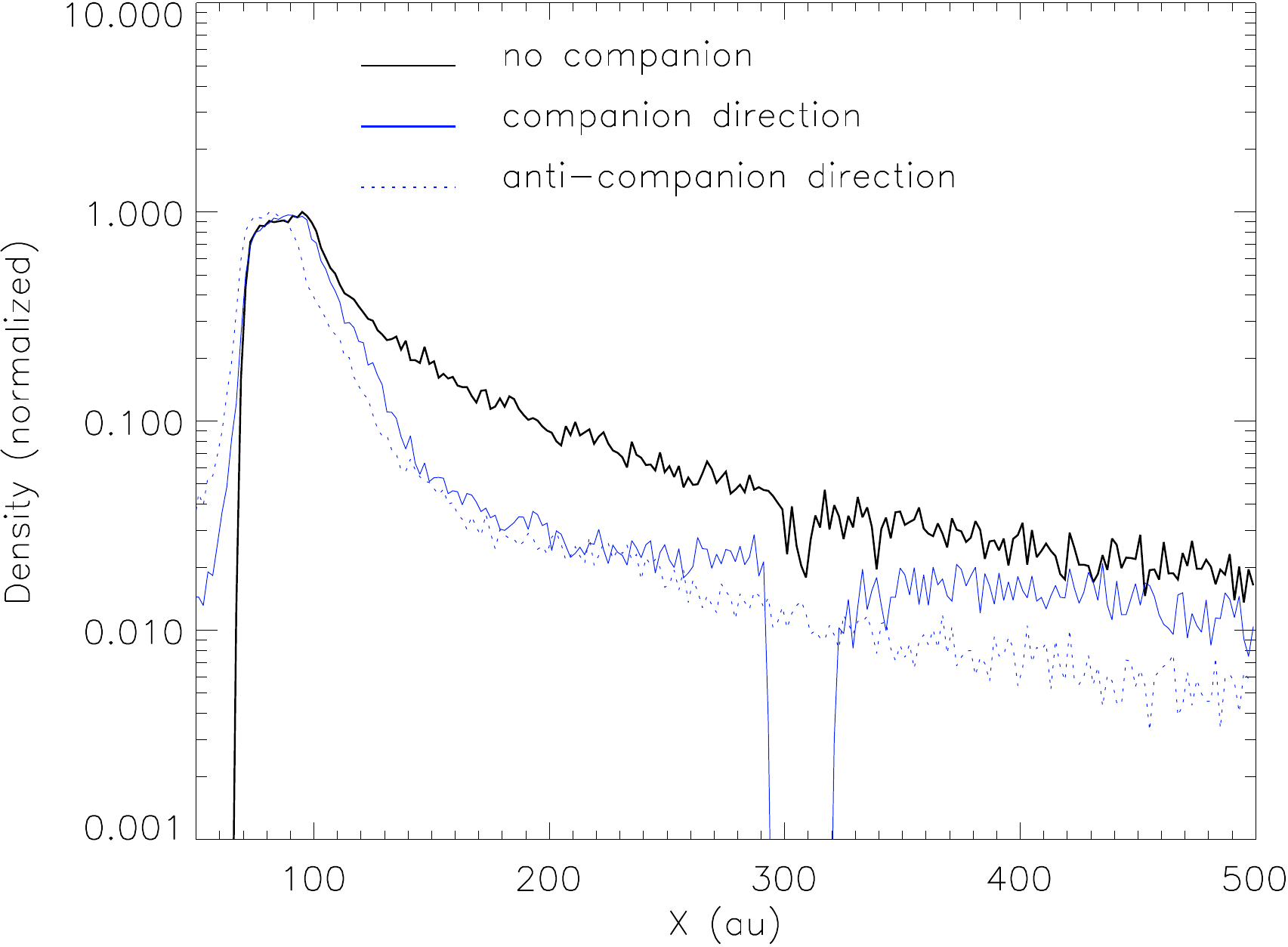}
\caption[]{Surface density radial profile for both the companion and anti-companion directions, as compared to the profile obtained with no companion. }
\label{densprof}
\end{figure}

\begin{figure*}
\makebox[\textwidth]{
\includegraphics[scale=0.45]{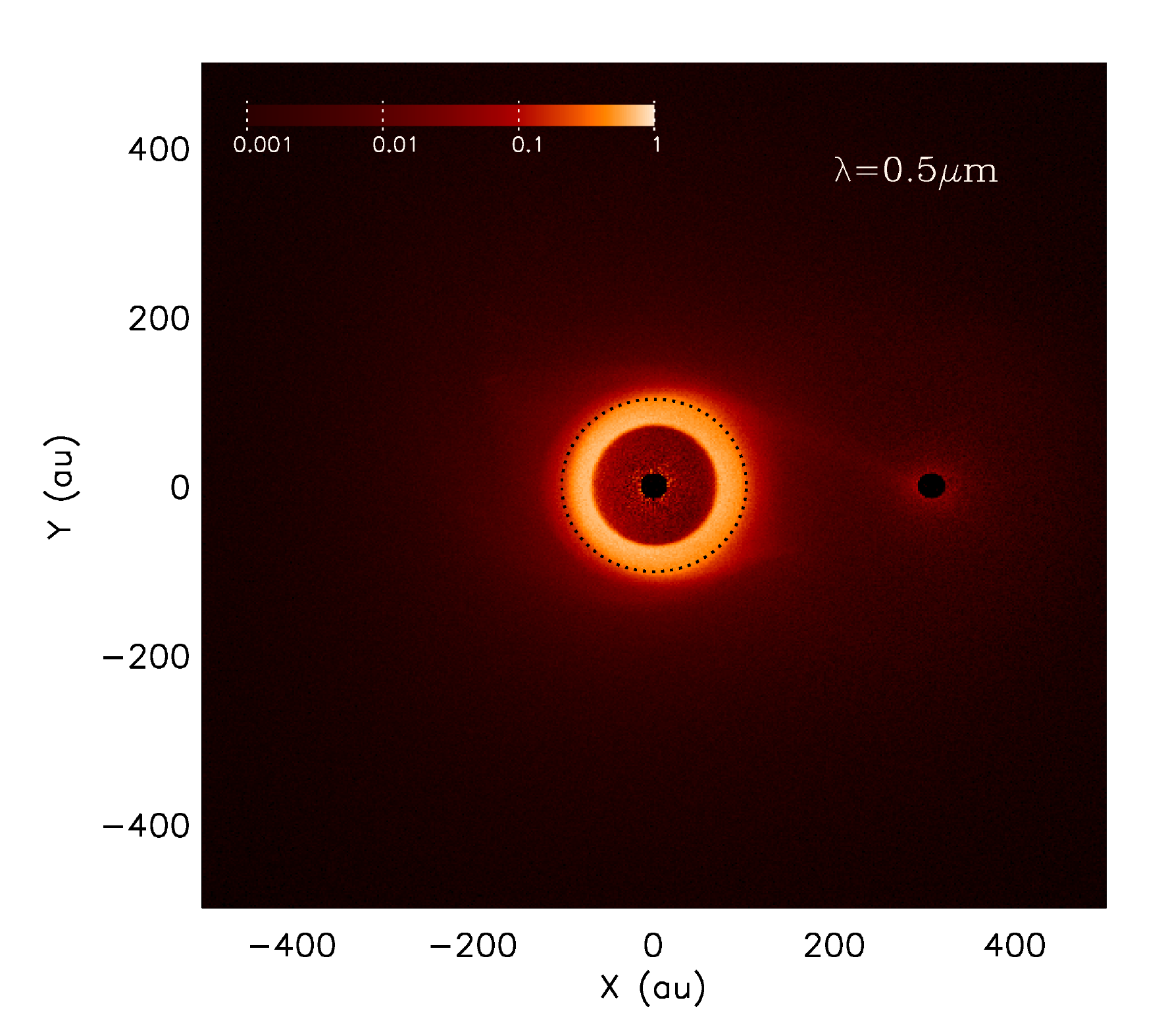}
\includegraphics[scale=0.45]{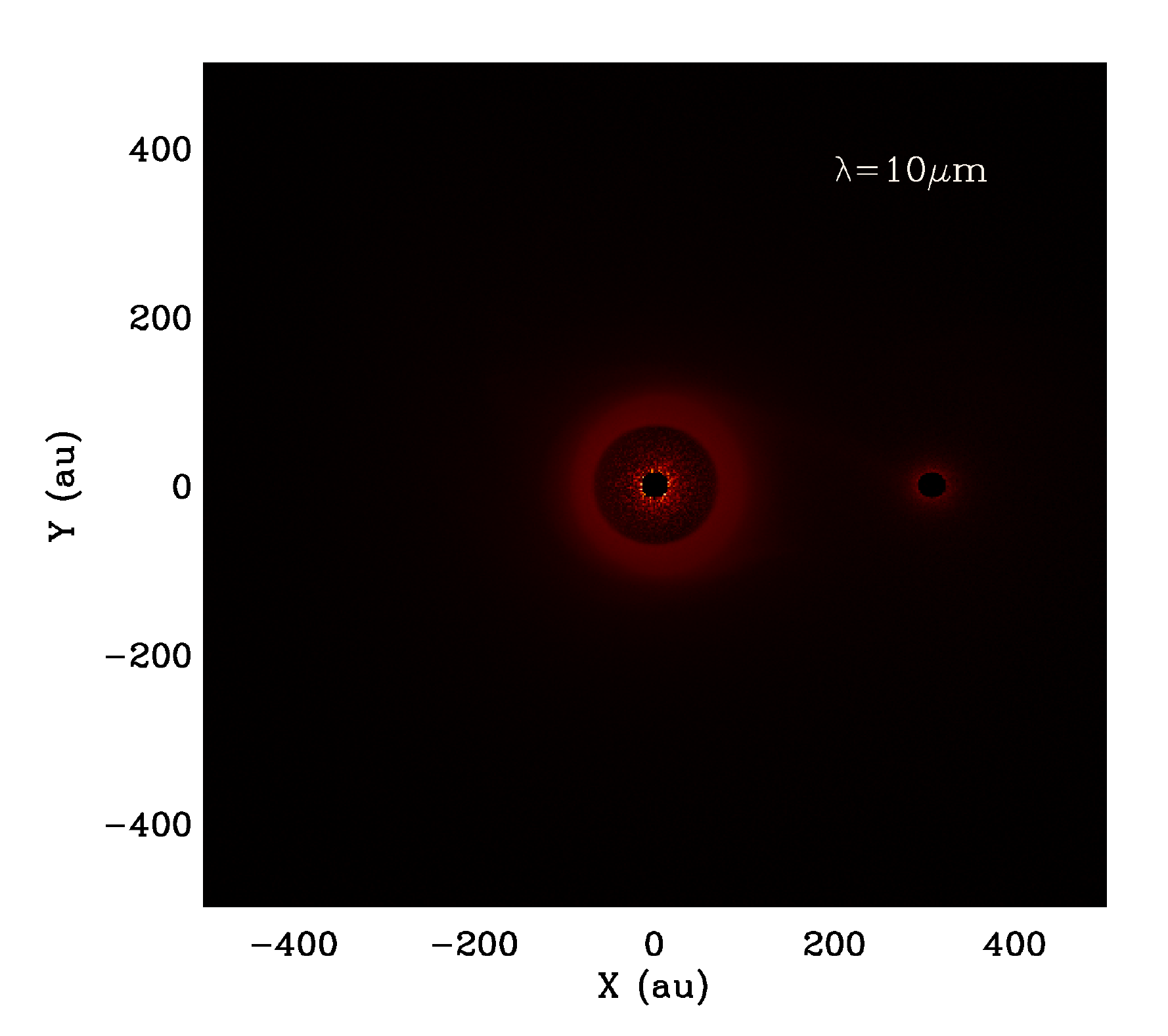}
}
\makebox[\textwidth]{
\includegraphics[scale=0.45]{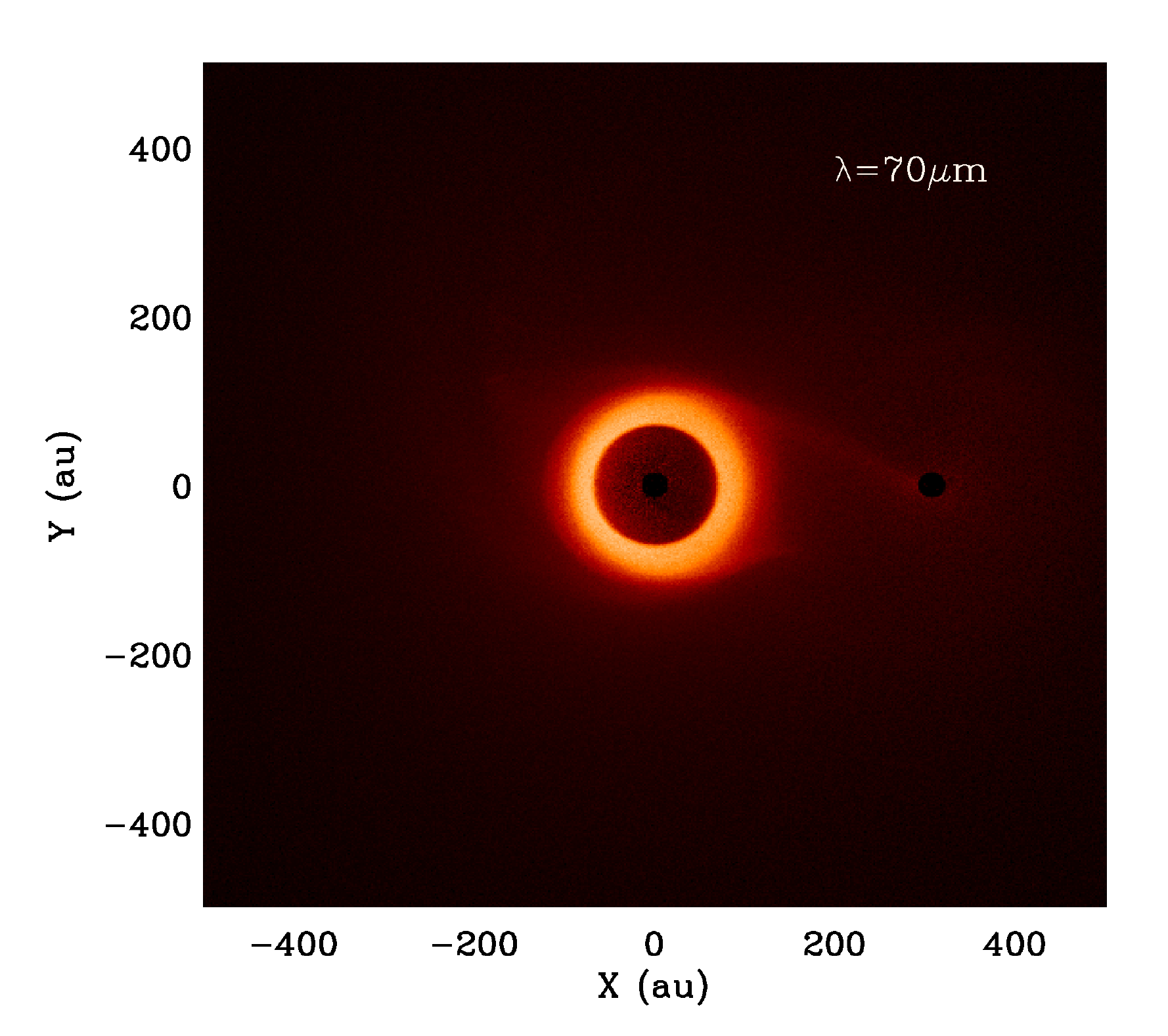}
\includegraphics[scale=0.45]{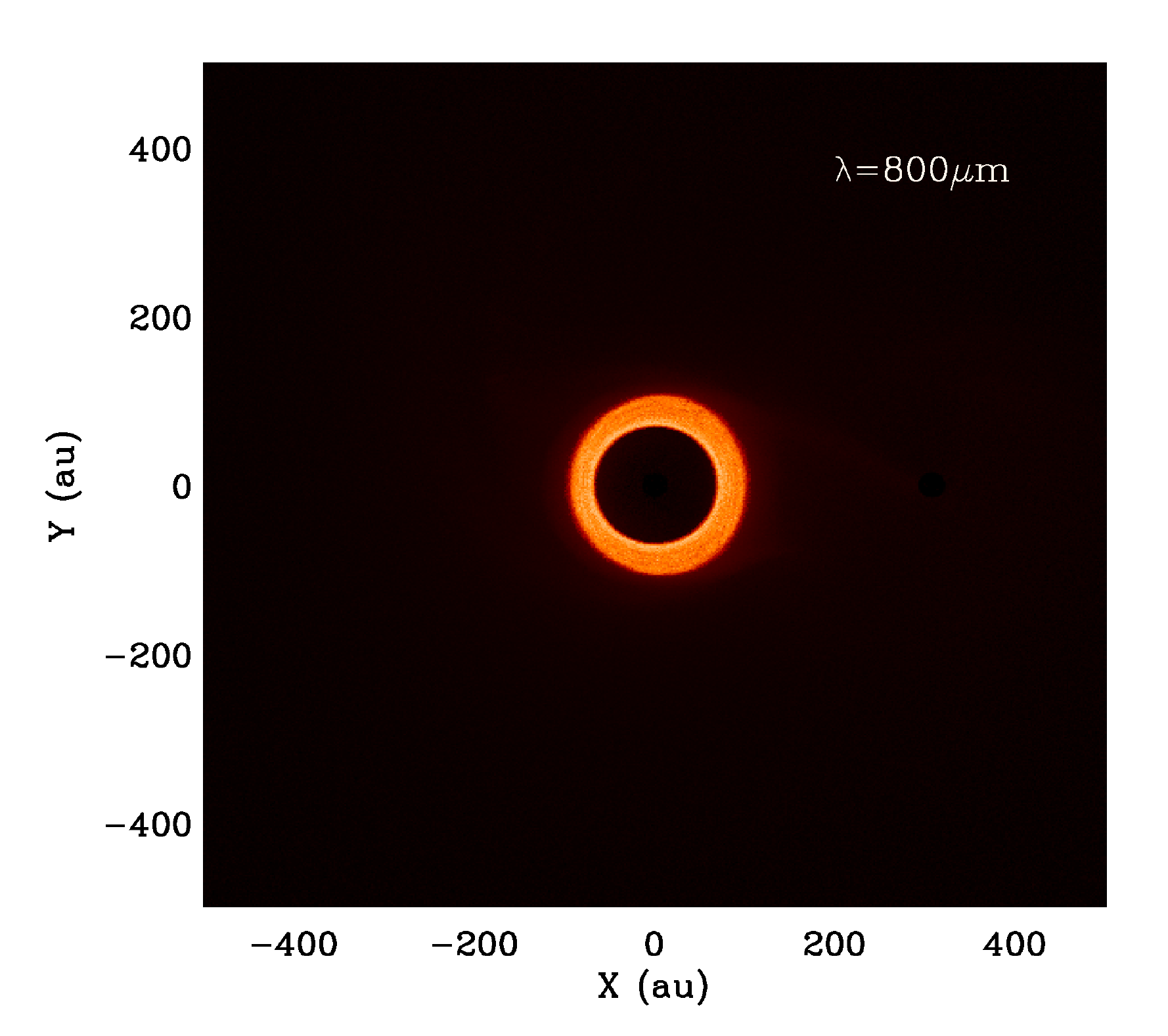}
}
\caption[]{Nominal case: Luminosity maps at four different wavelengths. The flux is re-normalized, in each image, to its peak value for the considered wavelength, and the relative colour-scale is the same in all 4 images. }
\label{e0im}
\end{figure*}

\begin{figure}
\includegraphics[scale=0.5]{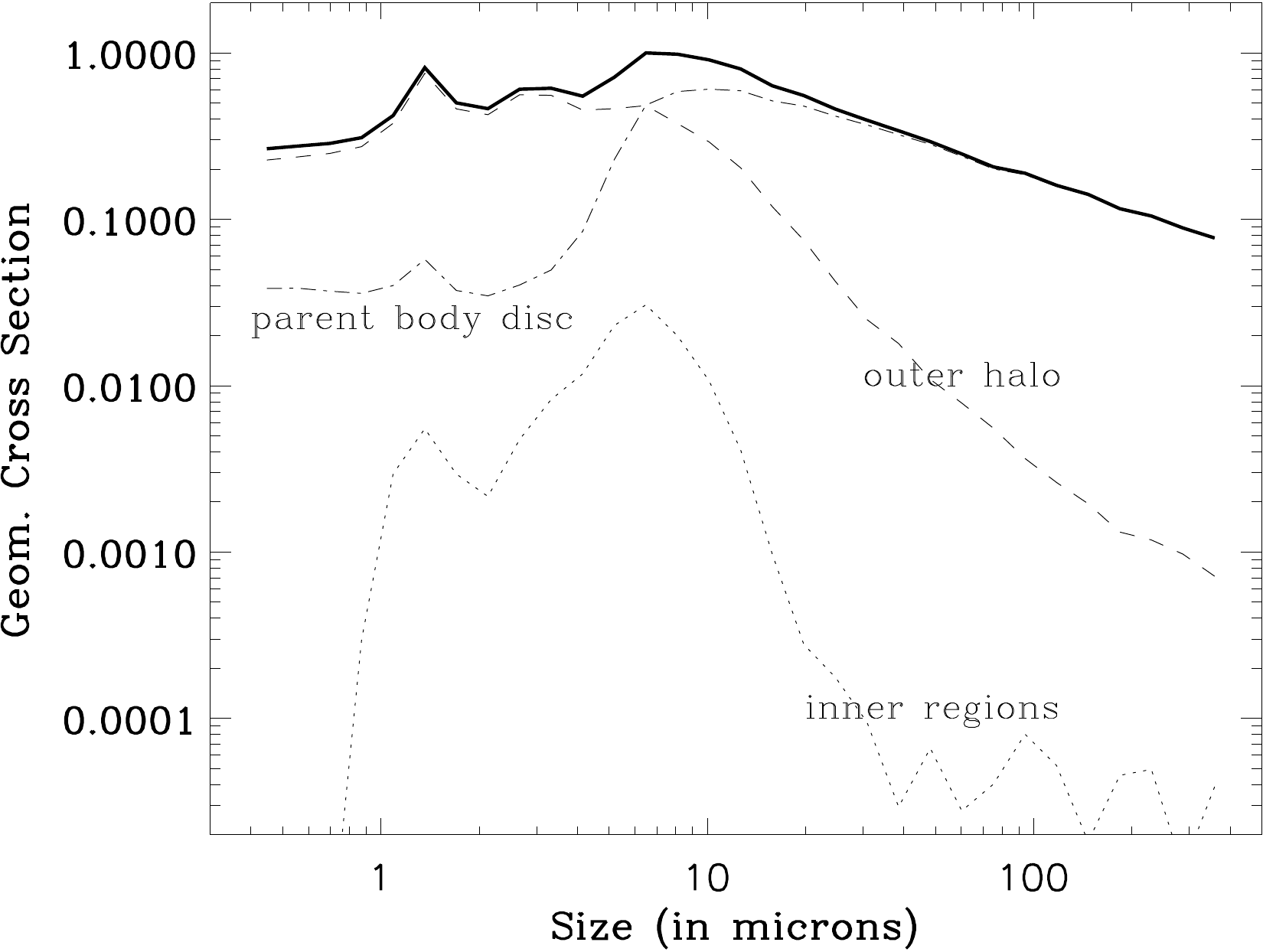}
\caption[]{Nominal case: Geometrical cross section per log grain-size for 3 different regions of the system. }
\label{psde0}
\end{figure}

\begin{figure*}[!h]
\makebox[\textwidth]{
\includegraphics[scale=0.5]{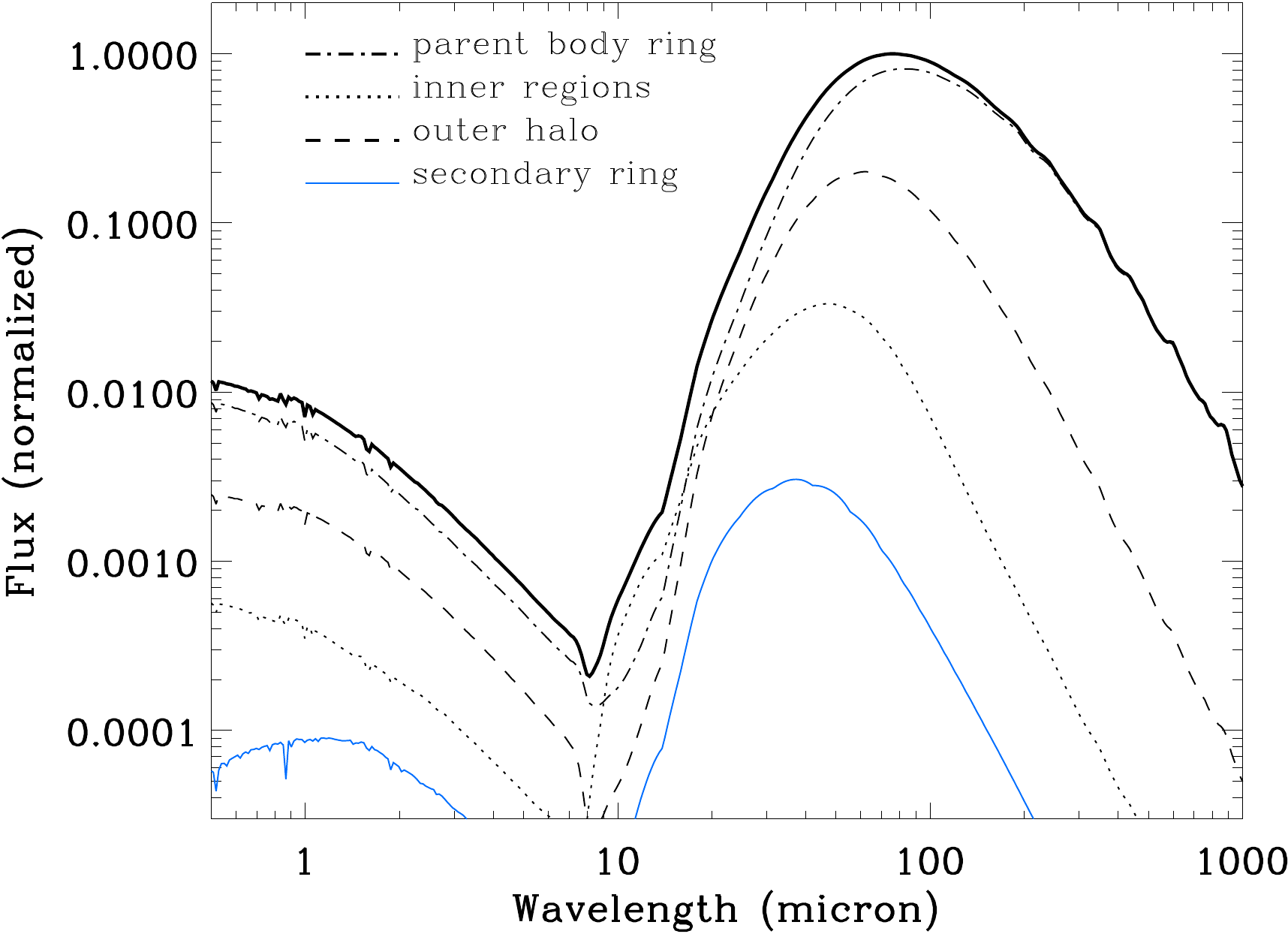}
\includegraphics[scale=0.5]{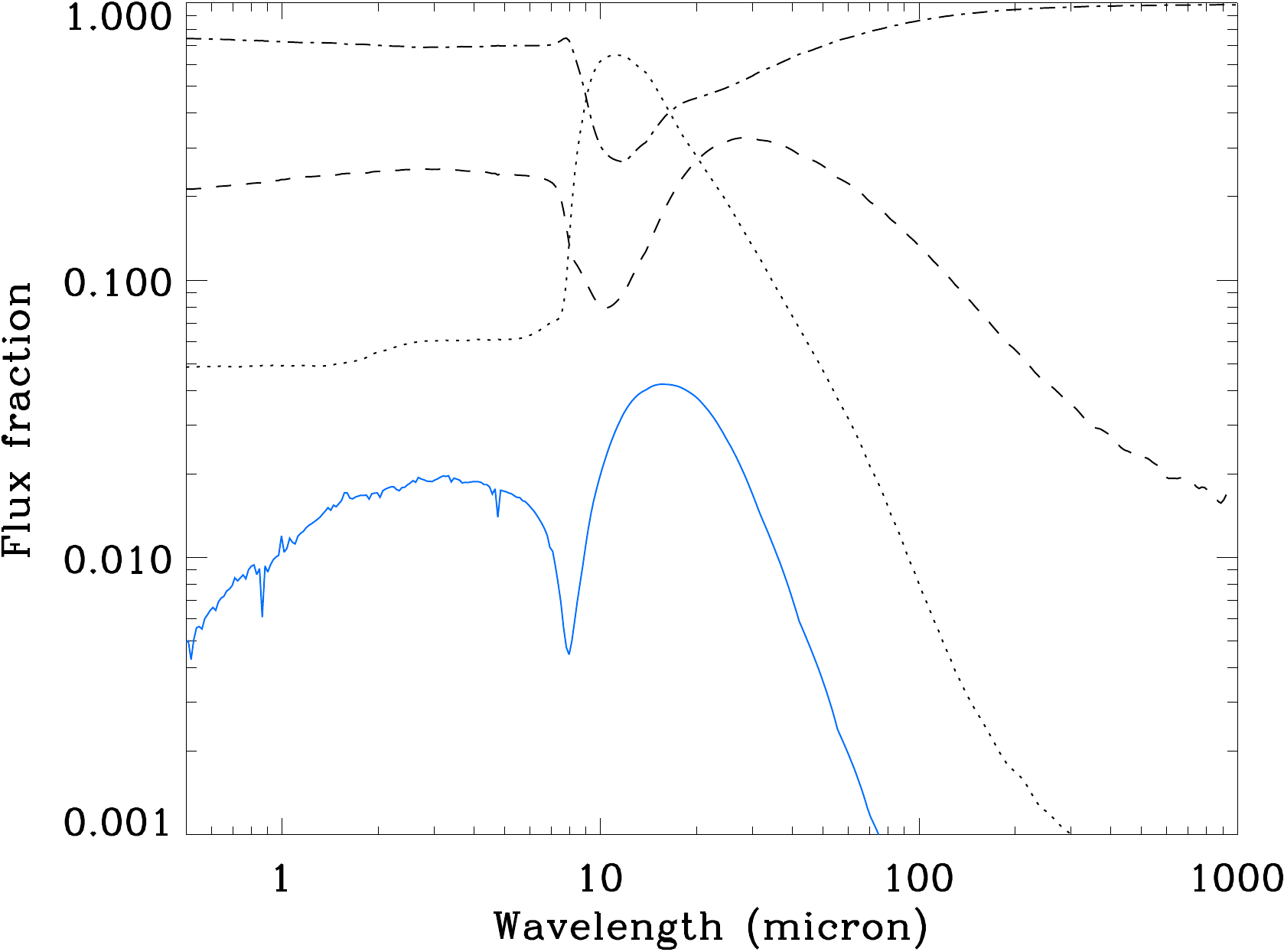}
}
\caption[]{Normalized Spectral Energy Distribution (SED) for the nominal case (left), and fraction of the flux coming from 4 different regions of the system as a function of wavelengths (right). The secondary ring corresponds to the flux coming from within 40\,au from the companion star.}
\label{sede0}
\end{figure*}

We use an updated version of the DyCoSS code, whose characteristics we briefly summarize here (for a more thorough description of DyCoSS, see TBO12). DyCoSS was, alongside the relatively similar CGA code \citep{stark09}, the first code able to handle, albeit in a simplified way, the \emph{coupled} effect of violent, fragment-producing collisions and dynamics (gravity and SPR). Its principle is to first run a classical $N$-body run for large parent bodies not affected by radiation pressure until a dynamical steady-state is reached. Each parent body is then used as a seed from which we generate a cloud of small debris affected by radiation pressure, and we run 10 different "debris" simulations for 10 different starting positions of the companion star on its orbit. In these simulations the density map of the parent bodies is used to estimate a local collisional destruction probability for each daughter particle. We then record, at regularly spaced time intervals, the position of each surviving particle, until all particles have been removed (either by dynamical perturbations or collisional destruction). Finally, we recombine together these 10 different simulations to produce collisional steady-state maps of the system for different positions of the companion star.

The main simplification in the DyCoSS code is that collisions are considered to be perfectly destructive: debris do not produce new debris but are removed after exceeding their estimated collisional lifetime. 
This constraint has been relaxed in the more sophisticated LIDT-DD code \citep{kral13}, based on collisionally interacting "super particles", which is the first (and so far only) fully self-consistent code coupling collisions and dynamics. However, the price to pay for this sophistication is that LIDT-DD is, in its current version, not very versatile. It has so far mostly been applied to study the aftermath of isolated violent collisional breakups \citep{kral15,theb18}, but it cannot easily handle cases with strong dynamical gradients such as an embedded planet or a binary companion. For such cases, DyCoSS remains better suited.

\subsection{Upgrades}

An important improvement regards DyCoSS' self-consistency with respect to collisional lifetime estimates $t_{\rm{coll}}$. Instead of computing $t_{\rm{coll}}$ from the density map of parent bodies, which can strongly depart from that of the debris they produce, we now use the final density map of the debris simulations. In order to do so, we proceed by iterations: we first run a preliminary set of "debris" simulations where $t_{\rm{coll}}$ is computed from the density map of parent bodies, but instead of stopping there (as we did in TBO12) we re-inject the obtained \emph{debris} density map to estimate collisional lifetimes in a new set of "debris" simulations. We then use the debris density map of these new simulations to compute $t_{\rm{coll}}$ in the next set of debris runs, and repeat this procedure until the final density maps converge within 1\% of the previous run, in an iterative approach that is close to that of the CGA code. 

Another upgrade concerns the spatial scale of the "debris" runs, which is no longer limited to the circumstellar region but now extends beyond the location of the companion star. We now also take into account the radiation pressure from this companion star.

Last but not least, we now couple our collisional code to the state-of-the-art GraTeR radiation transfer code \citep{auge99} to derive both realistic synthetic images at all wavelengths and global Spectral Energy Distributions (SEDs) of the integrated system. The GraTer package is used for both the contribution of the central star and that of the stellar companion.

\subsection{Setup}

We consider an initial disc of parent bodies extending from $r_{\rm{in}}=70\,$au to $r_{\rm{out}}=100$\,au of the primary star, which is the typical distance at which cold debris discs are  observed. The eccentricity $e_{\rm{b}}$ of the companion star's orbit is taken as a free parameter, and its semi-major $a_b$ is then chosen so that the outer edge of the initial parent body belt coincides with the outer limit $r_{\rm{crit}}$ for stable orbits around the primary star\footnote{The value of $a_b$ is obtained from a set of test runs.}, thus corresponding to the maximum possible effect of the binary on the disc. In other words, the companion star is as close as possible to the parent body belt without destroying it. For the primary star, we consider the nominal case of an A6V star of mass $M_1=1.7$ M$_{\odot}$, similar to the archetypal $\beta$-Pictoris case. Several values of the companion star's mass $M_2$ are explored between $0.25M_1$ and $M_1$, corresponding to M, K and A stars respectively, the $M_2=0.5M_1$ value being taken as a reference case. Note that, in order to clearly identify the role played by each parameter, we focus only on the evolution of a circumprimary ring and do not suppose the existence of a native secondary ring around the companion star.

As for the grain composition, we consider, for the sake of reducing the number of free parameters to handle, generic compact astrosilicates \citep{drai03}. The radiation pressure force $F_{\rm{PR}}$ is parameterized by the usual parameter $\beta_{\rm{(s)}}=0.5\,F_{\rm{PR}}/F_G$, where $F_G$ is the gravitational pull of the star. To compute $\beta_{\rm{(s)}}$ as a function of grain size and stellar luminosity from the primary star, we use the GraTeR package.  For astrosilicates and the A6V type we consider for the central star, the cut-off grain size below which grains are blown out on unbound orbits by the primary star is $s_{\rm{blow}}\sim2\,\mu$m. We consider a canonical grain size distribution in $s^{-3.5}$ starting from $s_\mathrm{min}=0.15\,\mu$m to $s_\mathrm{max}=1$ cm.

 As in previous studies with DyCoSS, we parameterize the collisional activity within the disc by the average vertical optical depth in the parent body ring <$\tau$>, which is a free parameter that is set before starting the runs.

For the sake of clarity, we consider a nominal set-up, corresponding to a circular binary ($e_{\rm{b}}=0$), a mass ratio $\mu=M_2/M_1=0.5$ (the companion is a K star) and a typical optical depth <$\tau$>$=2\times10^{-3}$ corresponding to the $\beta$-Pic debris disc. For this set-up the semi-major axis of the binary, chosen so that the stability limit coincides with the outer edge of the PB belt, is $a_b=310\,$au.
Each free parameter of the system ($e_{\rm{b}}$, $\mu$ and <$\tau$>) will then be explored individually from this reference configuration.

\begin{figure*}
\makebox[\textwidth]{
\includegraphics[scale=0.5]{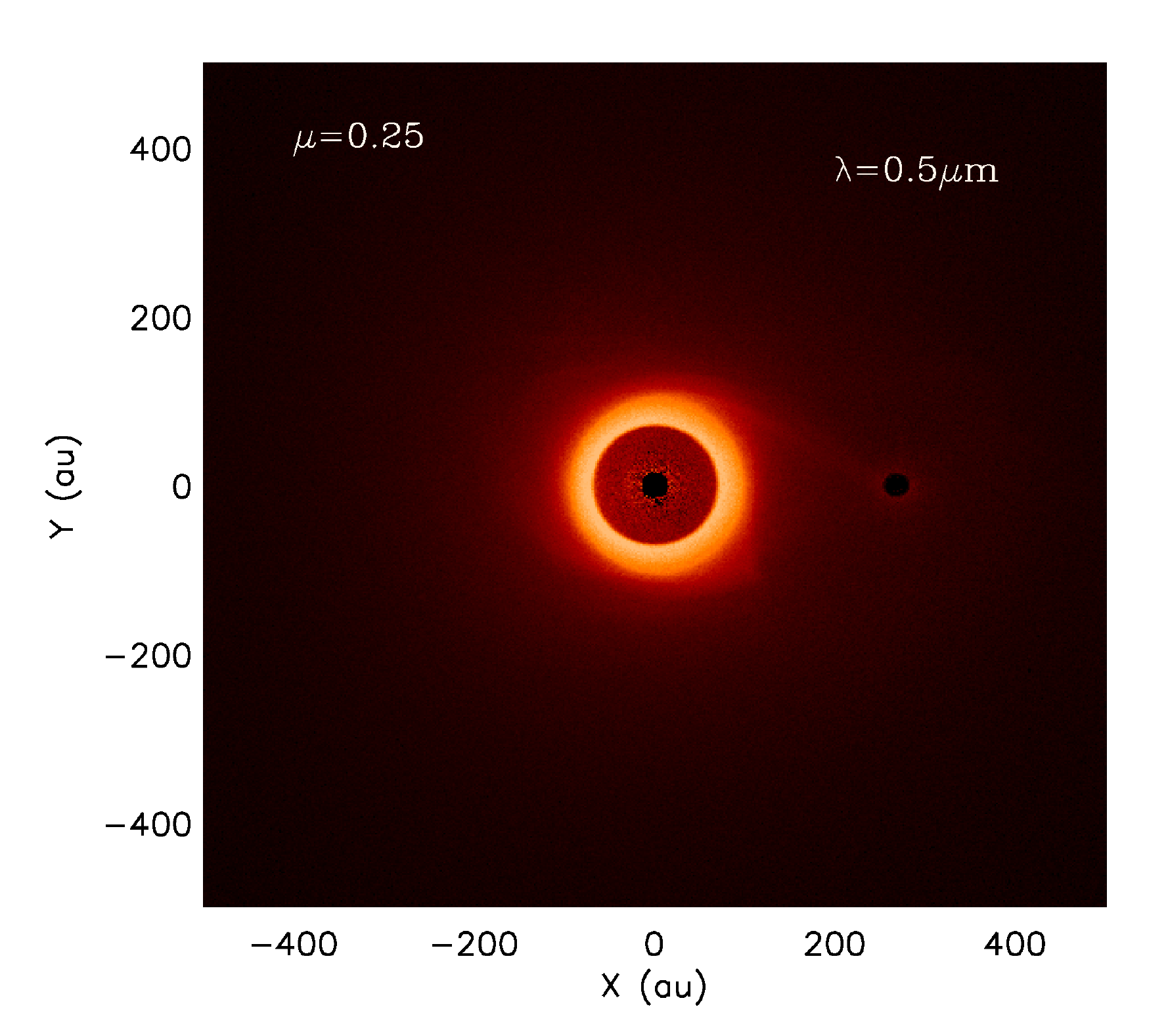}
\includegraphics[scale=0.5]{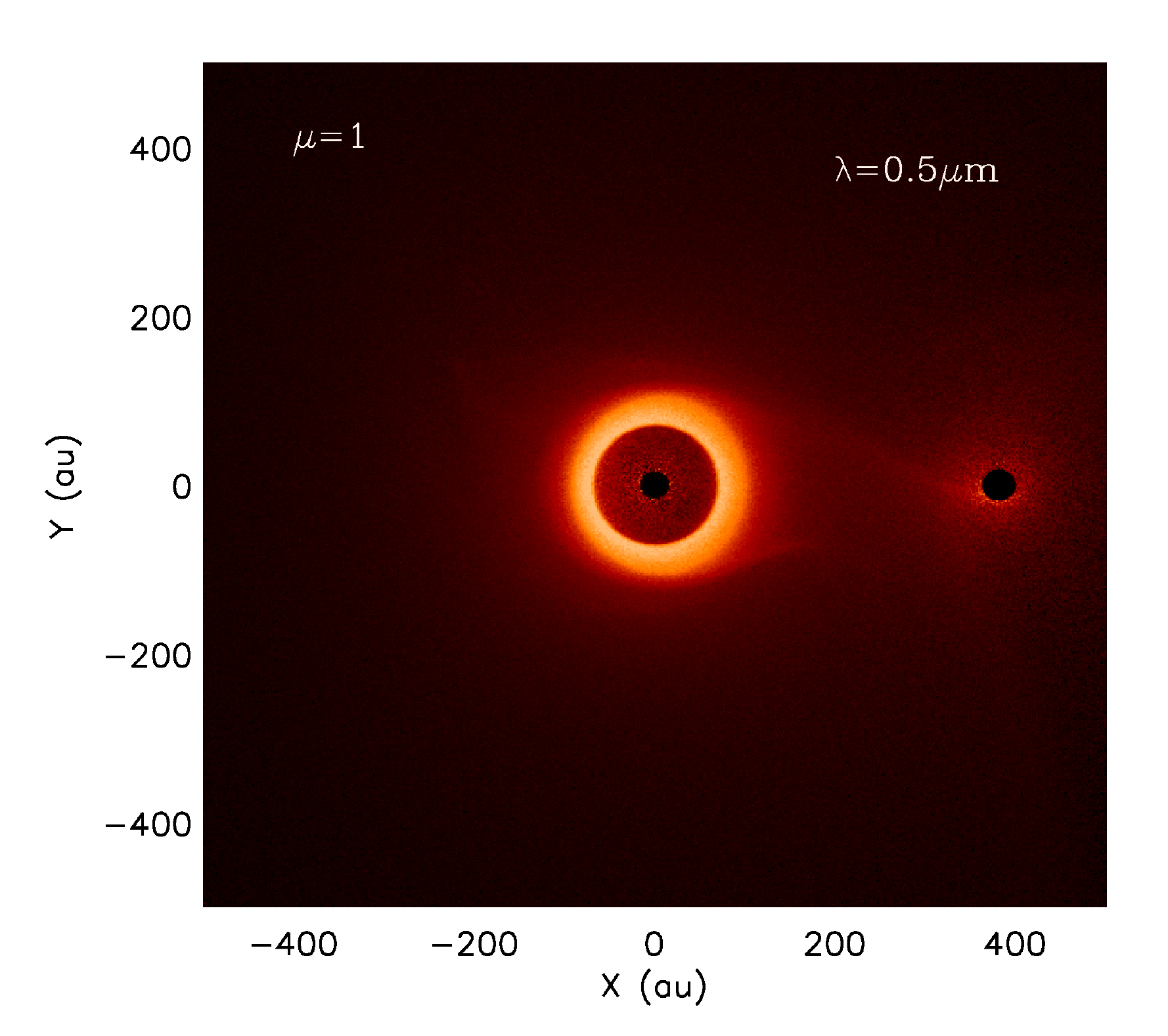}
}
\caption[]{Normalized scattered-light images for a low companion mass case with $\mu=0.25$ (left) and a massive companion case with $\mu=1$ (right) on circular orbits. Note that the companion location is automatically rescaled so that the dynamical truncation of the parent body ring falls at the same distance $r_{\rm{out}}$ of the central star. The colour scale is the same as in Fig.\ref{e0im}}
\label{surfmass}
\end{figure*}

\begin{figure}
\includegraphics[scale=0.5]{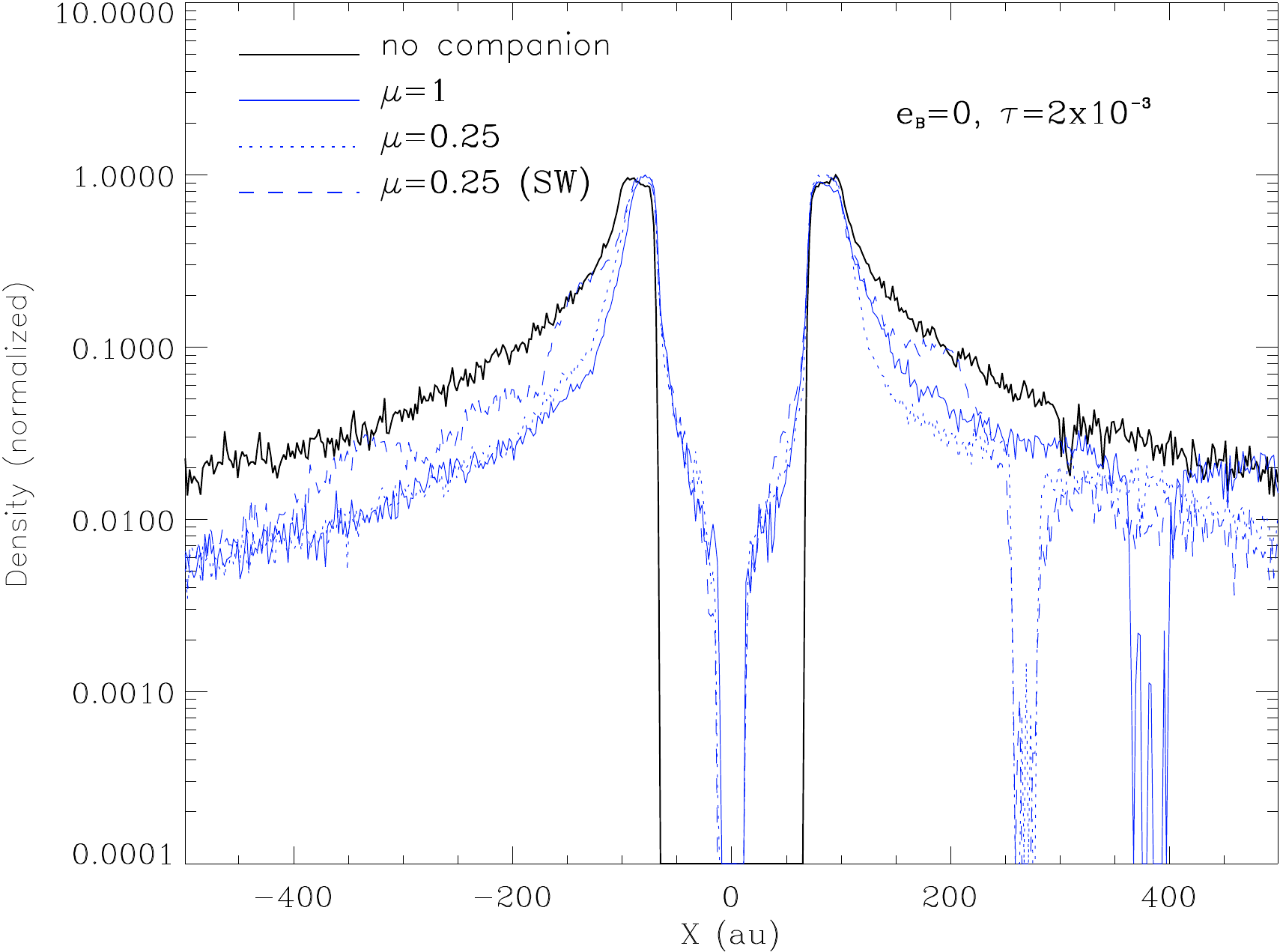}
\caption[]{Radial cut of the surface density along the axis passing by the companion (located on the left-hand side) when varying the mass of the companion star. The $\mu=0.25$\,(SW) case corresponds to a low-mass companion with a strong stellar wind. }
\label{cutcm}
\end{figure}

\section{Results}\label{results}

\subsection{Nominal case}\label{nomi}

Fig.~\ref{denscomp} presents a surface density map, at steady-state, for our nominal case ($e_{\rm{b}}=0$, $\mu=0.5$ and <$\tau$>$=2\times10^{-3}$). We confirm a crucial result of TBO12, which is that the steady collisional production of small, radiation-pressure affected grains is able to maintain a significant level of dust in the dynamically unstable region beyond $r_{\rm{out}}$. In these outer regions, dust density is still approximately 30\% of the level it would have in a case without stellar companion (Fig.~\ref{densprof}). Interestingly enough, the radial density profile in the anti-companion direction, after a significant drop just beyond the parent body ring, decreases as $\sim r^{-1.7}$ in the outer regions, a value that is remarkably close to the "canonical" slope in $r^{-1.5}$ obtained beyond the main ring of a companion-less system \citep[see black line in Fig.~\ref{densprof} and discussion in][]{stru06,theb08}. 
As for the parent body ring itself, it is moderately distorted by the companion star, with the ring being $\sim20\%$ wider in the direction of the companion than in the opposite one.

The larger spatial scale of the simulation allows us to identify, in the unstable region beyond $r_{\rm{out}}$, several structures that could only be hinted at in TBO12, in particular a spiral arm extending from the parent body ring all the way to the stellar companion and a much fainter secondary arm in the opposite direction. All these structures do not change shape with the stellar companion's orbital position and precess at the companion's angular velocity, i.e., slower than the local Keplerian velocity. 
We also identify a faint and wide circular ring around the companion star, of radius $\sim 120-140\,$au. This ring is made of unbound particles with sizes $s\leq s_{\rm{blow}}$, which should in principle be blown out by the central star's radiation pressure but are instead captured on stable orbits by the companion star. Contrary to appearances, these grains are located just inside the Roche lobe of the secondary. This is because their $\beta$ value is typically comprised between $\sim 0.7$ and $\sim 1$, so that they "feel" a primary star whose effective mass is reduced by at least a factor of $1/(1-\beta)\sim 3.3$, while the effective mass of the secondary star, whose radiation pressure is negligible, remains unchanged. As a consequence, for our nominal $\mu=0.5$ case, the secondary star has an effective mass at least 1.65 times higher than that of the primary, and thus a larger Roche lobe. For this mass ratio of 1.65 and our nominal value of $a_b=310\,$au, the approximate formula of \cite{eggl83} gives a Roche lobe of size $\sim 133\,$au, in good agreement with the location of the wide ring. The fraction of small grains trapped in this faint ring is, however, relatively limited:
for our nominal case we find that $\sim0.2$\% of all grains in the size range $0.5\,s_{\rm{blow}}\leq s\leq s_{\rm{blow}}$ produced in the parent body ring end up on a bound orbit around the companion. 

The luminosity maps in Fig.~\ref{e0im} reveal that the structures identified in the density distribution show up more prominently in two distinct wavelength domains: in scattered light and the mid-to-far IR ($\lambda \sim 70 \,\mu$m). This is particularly true of the wide spiral arm stretching from the parent body ring to the secondary star. The fact that the luminosity map at $\lambda_s=0.5\,\mu$m resemble the density map (weighted by the 1/$r^{2}$ dilution) is expected, because most grains in the simulation have a size $s\geq \lambda_s/2\pi\sim 0.1\,\mu$m, and thus a scattering efficiency close to 1. The fact that the $70\,\mu$m image presents a similar aspect might appear more counter intuitive, because most grains in the unstable $r>r_{\rm{out}}$ halo are in the $1\leq s \leq10\,\mu$m size range (Fig.~\ref{psde0}) and are thus mediocre emitters at this wavelength. But this characteristic is mitigated by the fact that most micron-sized grains in the 100-200\,au region have a temperature corresponding to a thermal emission peaking around 70$\,\mu$m. This peak at $\sim70\,\mu$m appears clearly when displaying the contribution of this outer halo to the system's integrated SED (Fig.~\ref{sede0}).  At even larger wavelengths, however, these micron-sized grains become too poor emitters to significantly contribute to the flux. This translates into a one order of magnitude drop of the halo's integrated contribution to the total SED beyond 70 $\mu$m (Fig.~\ref{sede0}b). As a consequence, with the same dynamical range as for the images at shorter wavelengths, the $\lambda= 800\,\rm{\mu}$m luminosity map only displays the belt of parent bodies in the $r\leq r_{\rm{out}}$ region. All structures in the unstable $r\geq r_{\rm{out}}$ domain are more than $10^3$ times dimmer than the PB belt at this wavelength.

The image at $\lambda=10\,\mu$m is radically different. Grains both in the parent body disc and the unstable halo are indeed much too cold to efficiently emit at these wavelengths, and the flux is instead dominated, at $\sim70$\%, by the small fraction of small grains that have been injected, on high-$e$ orbits, in the $r\leq r_{\rm{in}}$ region (Fig.~\ref{sede0}). This gives the illusion of an additional and more compact disc, that is brightest in the $\sim10-20\,$au region, where it surpasses the luminosity of the parent body belt. 

Note also that, at all wavelengths short of $\sim70\,\mu$m, a faint disc shows up around the secondary star. This disc is more compact than that around the primary and it is entirely made of small grains transferred from the primary disc's unstable halo \footnote{This compact secondary disc should not be confused with the tenuous wide secondary ring identified on the density maps. This wide ring does not show up on the present luminosity maps because it scatters too little light from both stars and is made of $s\leq s_{blow}$ grains that are too poor emitters in thermal light.}. Its brightness comes from the secondary star's radiation, even though its luminosity remains relatively low. In scattered light, we estimate that its maximum brightness per au$^2$ is $\sim 1.3$\% that of the main disc around the primary. 
Its IR fractional luminosity is $\sim 6\times10^{-6}$ that of the primary star, but $\sim10^{-4}$ that of the secondary.

\subsection{Role of the binary mass ratio}

The low mass ($\mu=0.25$) and massive ($\mu=1$) companion cases do not lead to images that drastically differ from our nominal ($\mu=0.5$) case in scattered light (Fig.~\ref{surfmass}). This is confirmed by the radial cuts of the surface density (Fig.~\ref{cutcm}), showing that the profiles are relatively similar for all companion masses in the $0.25\leq \mu \leq 1$ range, except for the region close to the secondary star itself. This might appear counter intuitive but is partly due to the fact that, for each case, the companion's position is rescaled so that the dynamical truncation of the parent body ring falls at the same distance $r_{\rm{out}}$ of the central star (meaning that, for instance, a more massive secondary star will be located further out). In this respect, what our results show is that, \textit{if} a parent body disc is truncated by an outer stellar companion, then the steady-state structure of the debris disc linked to these parent bodies does not strongly depend on the companion mass (all other parameters been equal). The only significant difference concerns the compact secondary disc of particles temporarily or permanently trapped around the secondary, which is significantly brighter, for the massive $\mu=1$ case, than for the nominal $\mu=0.5$ run. In scattered light, this secondary disc's peak brightness per au$^2$ reaches, for the $\mu=1$ case, $\sim10\%$ that of the primary ring, although its detectability on resolved images would be more challenging, given that its most luminous parts are within 10-20\,au of the companion star (see section \ref{secring}).
As for its fractional luminosity, it is approximately $5\times10^{-5}$ that of both the primary and secondary stars.

\subsection{Effect of an intense stellar wind from the companion}

\begin{figure}
\includegraphics[scale=0.5]{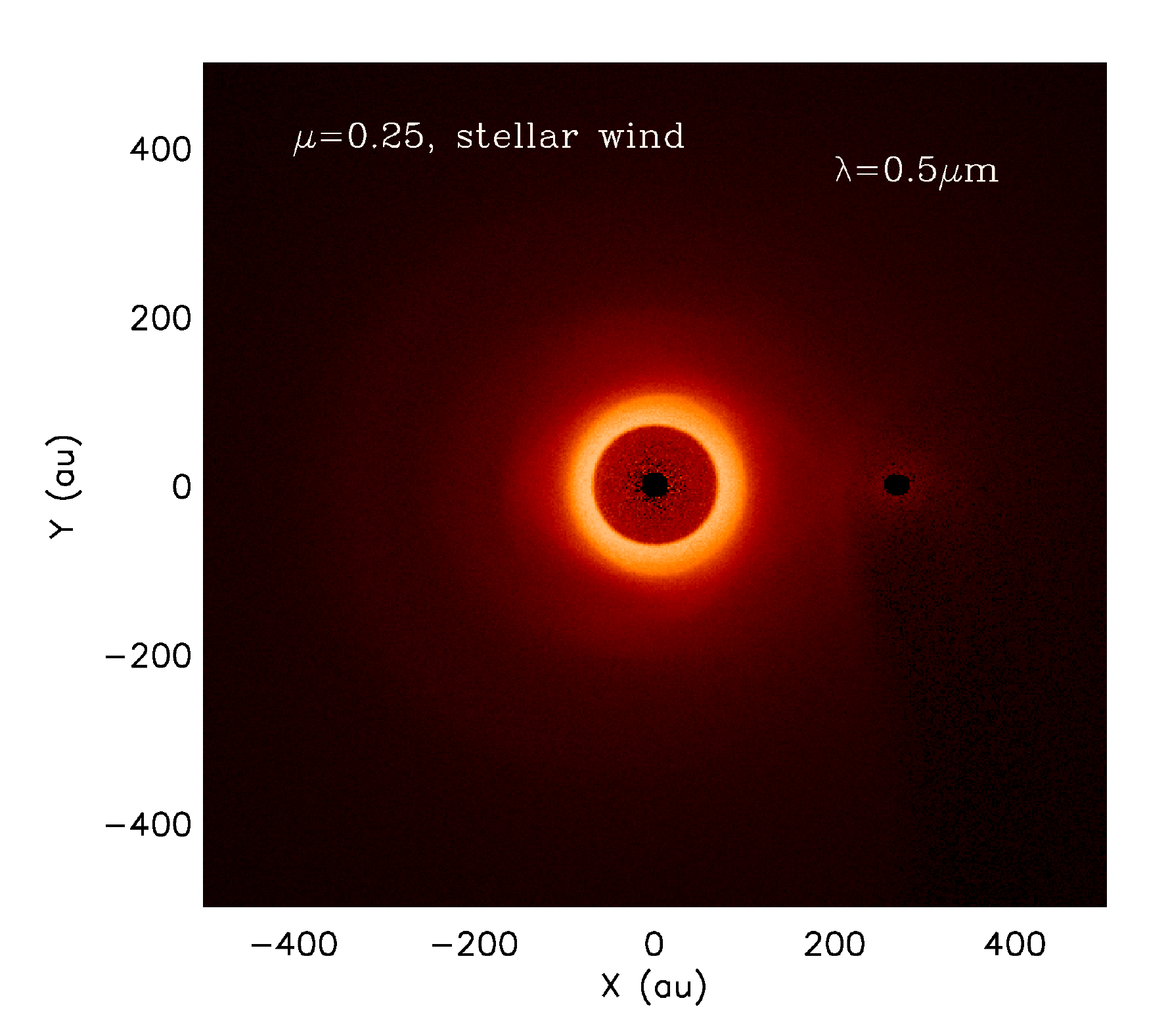}
\caption[]{Normalized scattered-light images for a low companion mass case with $\mu=0.25$, for which we assume a strong stellar wind that blows out all grains $s\leq s_{\rm{SW}}=6\,\rm{\mu}$m. The colour scale is the same as in Fig.\ref{e0im}}
\label{surfmasspr}
\end{figure}

We also explore, for the low-mass case $\mu=0.25$, the possible influence of the strong stellar wind that is potentially produced by such M stars \citep{seze2017}. Following previous studies of this mechanism, we assume that the stellar wind force on a dust grain follows the same radial dependence in $r^{-2}$ and the same size dependence in $s^{-1}$ as the radiation pressure force. It can then be parameterized by the cutoff size $s_{\rm{SW}}$ below which particles are blown out by the wind. In order to clearly identify its potential effect we consider a relatively strong wind \footnote{For the sake of simplicity, the drag component of this wind force is not taken into account. However, test simulations have shown that it did not affect our results on the timescales we are studying here.} with $\beta_{\rm{SW}}=6\,\rm{\mu}$m. The scattered-light synthetic image in Fig.~\ref{surfmasspr} shows several interesting features. The first one is a much brighter halo in the "forbidden" $r\geq r_{\rm{crit}}$ region, where the surface density is roughly twice what it is without stellar wind (Fig.~\ref{cutcm}). This is mainly because most micron-sized grains feel a repulsive force from the secondary that prevents them from easily leaving the halo around the primary. Another consequence of this repulsive force is to create an accumulation of small grains at the point where the stellar wind slows down or even reverses their outward motion. This location corresponds to the arch-like structure that is seen two-third of the way between the two stars. It is even clearer on the radial cut, where it shows up as a bump around 150-180\,au in the direction of the secondary (Fig.~\ref{cutcm}).

\subsection{Role of orbital eccentricity}

\begin{figure*}
\makebox[\textwidth]{
\includegraphics[scale=0.5]{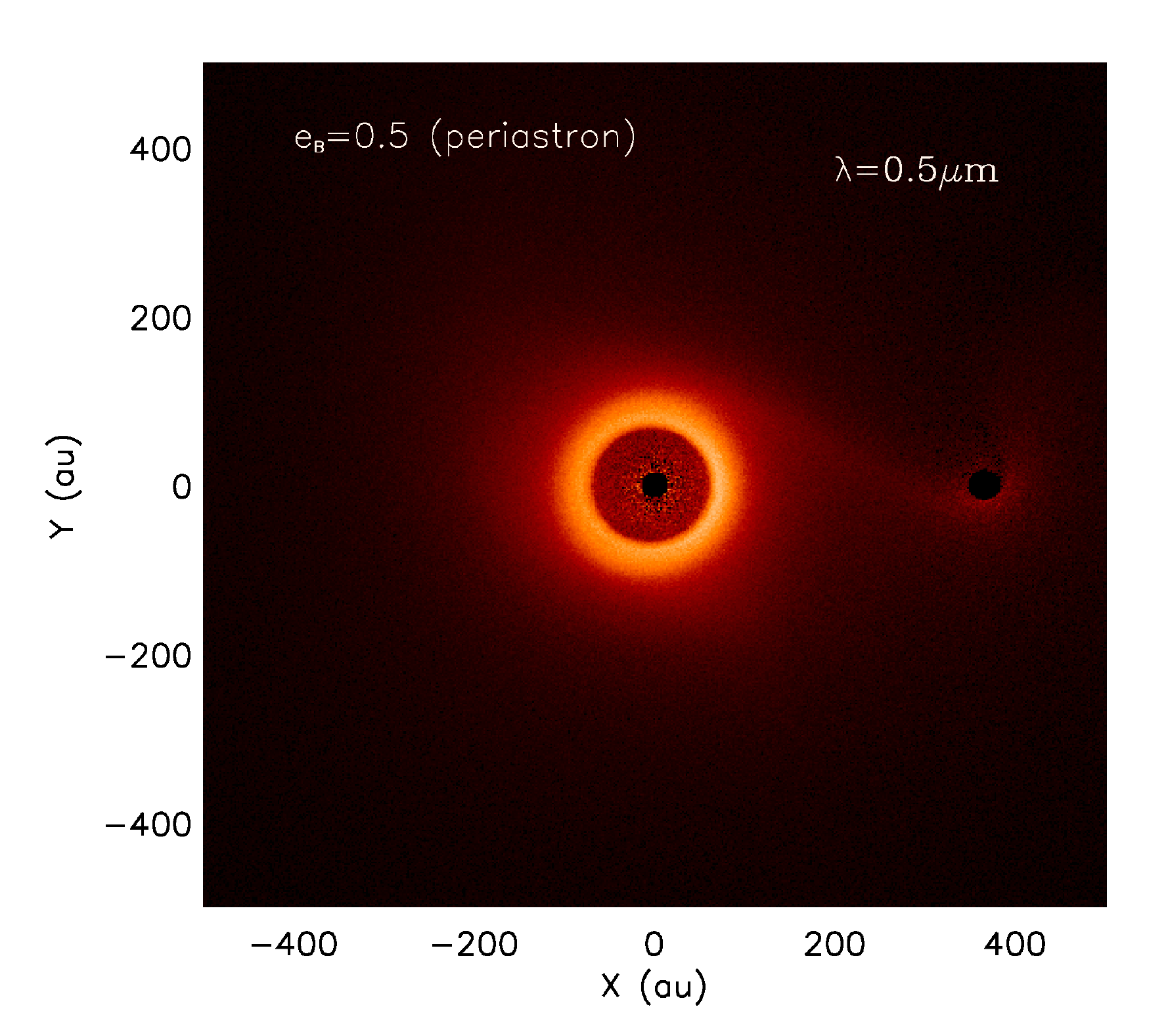}
\includegraphics[scale=0.5]{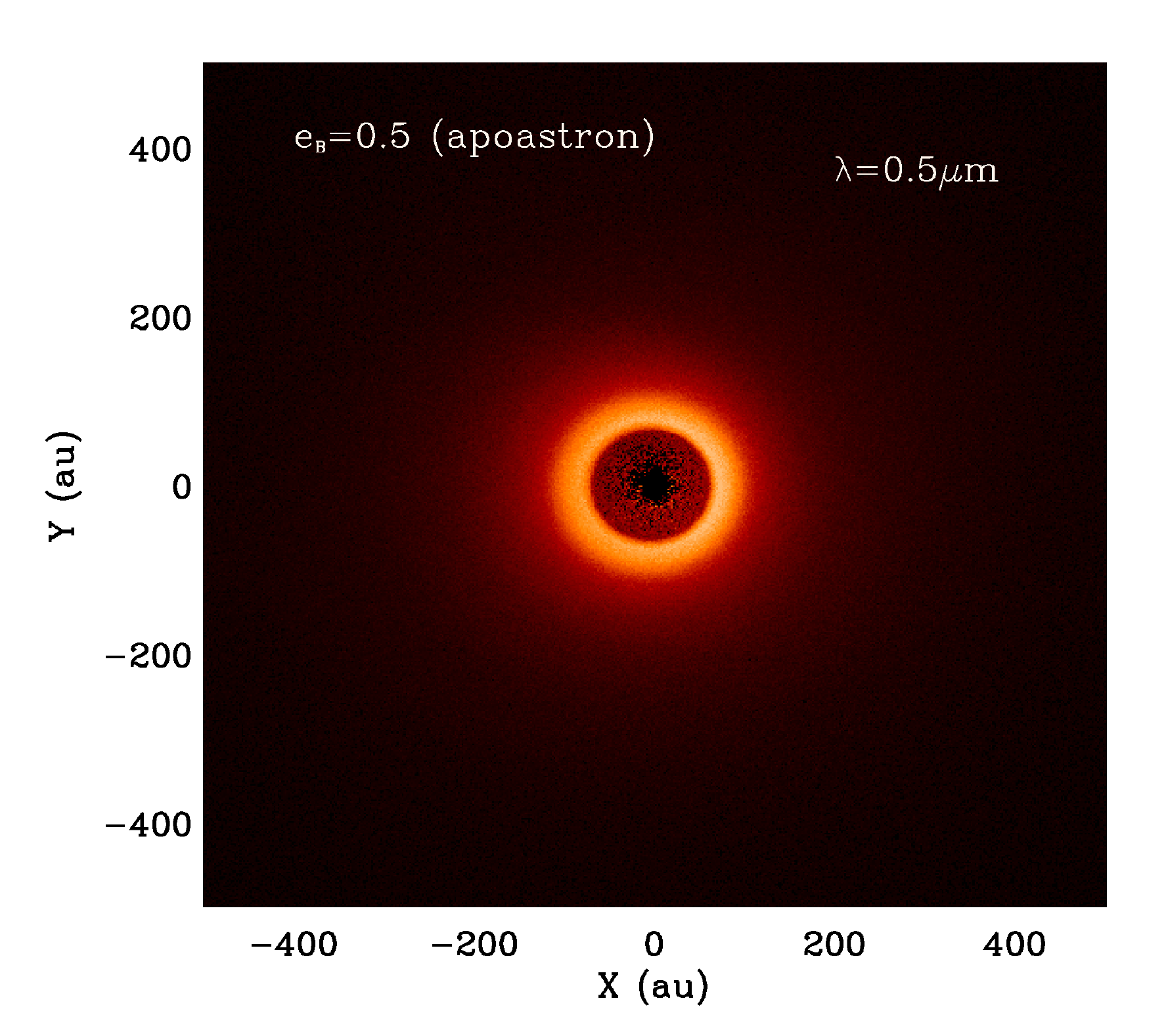}
}
\caption[]{Normalized scattered-light images for an eccentric companion with $e_b=0.5$, at periastron (left), and at apoastron (right). Note that this apoastron distance is 1095\,au, so that the companion is located outside of the field of the view in the right panel. The colour scale is the same as in Fig.\ref{e0im}. An animated version of this figure, showing the disc's evolution over one binary orbit, can be found at \url{https://lesia.obspm.fr/perso/philippe-thebault/anim-e05.gif}}
\label{mapeb}
\end{figure*}


Increasing the companion's eccentricity (while still adapting its location in order to truncate the parent body belt at $r_{\rm{out}}$) has a pronounced effect. As already noticed in TBO12, for high $e_{\rm{b}}$ the disc's intrinsic aspect varies with the companion's orbital position instead of keeping the same shape and precessing with the companion. For these high $e_{\rm{b}}$ values, the system displays pronounced structures, such as the single arm linking the PB belt to the companion, only at periastron passages of the companion and in their immediate aftermath, while adopting the shape of a relatively stable elongated disc the rest of the time (Fig.~\ref{mapeb}). 
The secondary disc around the companion is also only visible close to periastron passages and becomes highly asymmetric: it is much brighter on the trailing side of the companion's orbit than on the opposite side (Fig.~\ref{mapeb}a).

\subsection{Role of the disc's surface density}

\begin{figure}
\includegraphics[scale=0.5]{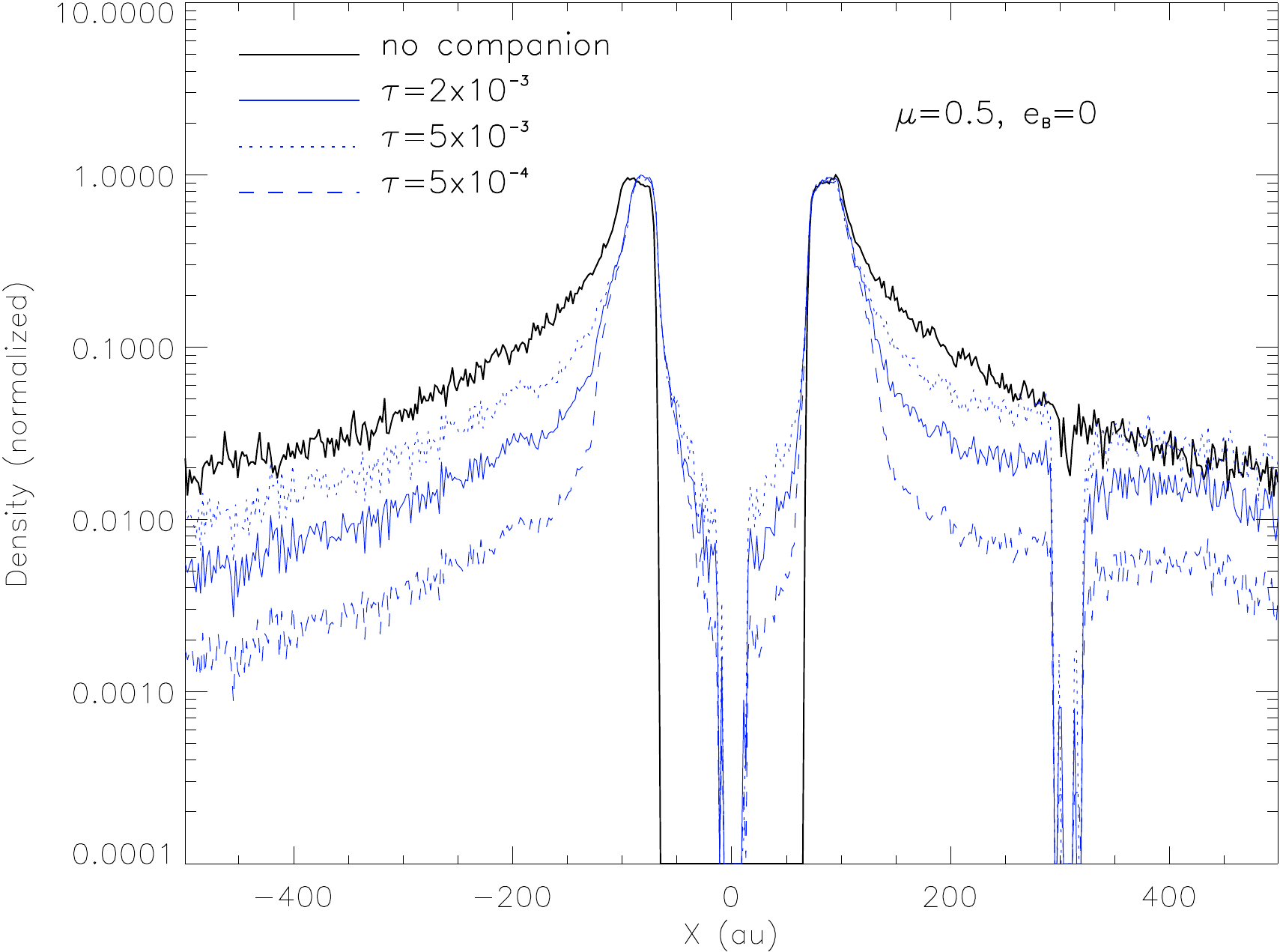}
\caption[]{Radial cut of the surface density along the axis passing by the companion for different values of the parent body's average optical depth, that is for different levels of collisional activity.}
\label{cutcd}
\end{figure}

A parameter that has a strong influence on the disc's structure is the average optical depth <$\tau$>, which sets the level of collisional activity within the parent body ring. If the slightly elongated shape of the PB ring, which is imposed by the dynamical effect of the companion alone, remains the same regardless of the value of $\tau$, the level of dustiness beyond $r_{\rm out}$ strongly varies with this parameter (Fig.~\ref{cutcd}).
This is fully expected because the amount of dust present in these dynamically unstable regions depends on the competing effect between the rate at which this dust is collisionally produced in the PB ring and the rate at which it is removed by the companion's perturbations. As a consequence, for denser discs ($\tau=5\times10^{-3}$) the system's steady-state logically approaches that of a companion-free case, whereas it approaches the pure dynamical case, with a sharp cutoff at $r_{\rm crit}$, for lower $\tau$ values (Fig.~\ref{cutcd}).

\section{Discussion} \label{discu}

The primary goal of this study is to investigate the signatures binarity can leave on a circumprimary debris ring, both on resolved images (primarily in scattered light) and in disc-integrated SEDs, and to assess to what extent these signatures can be unambiguously attributed to binarity and compared to observed systems. We stress that, in our parameter exploration, the companion star is always located as close as possible to the circumprimary belt of parent bodies without disrupting it. In other words, the outer limit of the PB belt corresponds to the limit for orbital stability $r_{\rm{crit}}$ around the central star, which is equivalent to studying the maximum possible effect of a companion star. This corresponds to a companion whose closest approach is at typically 3 to 4 times $r_{\rm{crit}}$.

\subsection{Signature of binarity on resolved images}\label{resolved}

\subsubsection{general trends}\label{trends}

Our numerical investigation has confirmed that, provided that the disc is sufficiently collisionnaly active (with $\tau$ typically larger than a few $10^{-4}$), there is always a halo of small particles beyond the stability limit $r_{\rm{crit}}$, created by the concurring effect of collisional production of fragments in the PB belt, radiation pressure from the central star and dynamical removal by the companion. Depending on the system's configuration (binary orbit and mass, disc density), this halo has a surface density comprised between $\sim10$ and $\sim50$\% that of the "natural" small grain halos that are routinely observed beyond PB rings in a non-binary environment \citep{theb08}. As such, it should be clearly visible on resolved images of bright discs, at least up to the mid-IR, provided that the companion star can be hidden with a coronagraph or be out of the field of view.

Companion perturbations also create a second halo, this time in the inner regions short of $r_{\rm{in}}$. While this inner halo is less dense than the outer one (Fig.~\ref{densprof}), it is closer to the star, increasing its brightness both in scattered light and thermal emission, and it is even dominating the system's total flux around $10\,\mu$m. However, given that the $\lambda\sim10-20\,\mu$m domain is precisely the one where the system is the dimmest (Fig.~\ref{sede0}), its observation might be very challenging.

It is in the outer dynamically "forbidden" halo that most binary-induced spatial features are to be found. The most striking one is a spiral-like pattern that stretches all the way from the parent body belt to the companion star. For binaries on low-eccentricity orbits, this arm is present for the whole binary orbit and precesses at the companion's angular velocity. Its brightness relative to the main PB belt is highest in two wavelengths domain: in scattered light and at $\lambda \sim 70\,\mu$m.
In scattered light, its luminosity is roughly 50\% higher than that of the background halo, making its detection at least as feasible as that of the halo itself.
Note that there is also a second spiral arm in the anti-companion direction, but it is much fainter than the primary one. This result is reminiscent of
results obtained in simulations of gaseous proto-planetary discs in binaries \citep[e.g.][]{dong2016}, which also display such a double spiral structure with the arm pointing at the companion being brighter than the secondary arm. There are, however, important differences, notably that, for gaseous discs, the main spiral arm is truncated and never extends to the companion star. These differences are expected because of different physical processes at play for the two cases: the evolution of the gaseous disc is driven by the coupling between binary perturbations and disc viscosity whereas for a debris disc it is the coupling between binary perturbations and stellar radiation pressure that is crucial.

For high-eccentricity binaries, the arm is only clearly present close to periastron passages, while the rest of the time the halo is almost featureless and axisymmetric. These high-$e_{\rm{b}}$ cases are the ones for which the companion has the strongest effect on the parent body ring itself, which becomes slightly eccentric itself, leading to the well-known pericentre glow effect \footnote{This effect is due to the fact that grains at pericentre appear brighter than at apocentre both in the optical, because they scatter more stellar light, and in thermal emission, because they are hotter.}, making the ring side closer to the primary star brighter than the opposite one, as appears clearly on Fig.~\ref{mapeb}. 

 For all cases (except for high-e binaries where it appears only close to periastron passages), a faint and compact secondary disc shows up around the companion star in scattered light up to the mid-IR. Note that this disc is not native but made of dust transferred from the unstable circumprimary halo. As a consequence, it is mostly made of small grains, with a strong depletion of particles larger than $\sim10\,\mu$m (Fig.~\ref{psde0}) that sets it apart from a native circumsecondary ring. This secondary ring is brightest for the case of a massive companion ($\mu=1$).
 
Apart from the secondary ring's brightness, the mass of the companion has a relatively limited effect on the system's structure, mainly because the companion's semi-major axis is always rescaled in order for $r_{\rm{crit}}$ to fall at the outer edge $r_{\rm{out}}$ of the parent body belt (here at 100 au), meaning that more massive companions are located further away from the PB ring. The only clearly outlying result is obtained for a low-mass companion that produces a very strong stellar wind. In this case micron-sized grains stay much longer in the halo, which becomes almost as bright as in a companion-free case. At the limit where SW effects are of the same order of magnitude as radiation pressure from the primary, an arc-like structure forms, which is brightest in scattered light. This arc structure also cuts off the arm-like feature that connects the PB ring to the secondary for all other cases.

\subsubsection{comparison to observed discs}

Of the 38 known confirmed disc-hosting binaries identified by \cite{yelv19}, only 7 have so far had their discs been unambiguously resolved\footnote{See the Jena database (\url{https://www.astro.uni-jena.de/index.php/theory/catalog-of-resolved-debris-disks.html})}, in scattered light or mid-IR images: HD 20320 \citep{boot13}, HD 102647 \citep{matthews2010}, HD 165908 \citep{kenn12}, HD 172555 \citep{smit12, engl18}, HD 181296 \citep{smit09}, HD 207129 \citep{kris10} and Fomalhaut \citep{holl98,kala05}. Of these 7 systems, only Fomalhaut has been imaged at a resolution and with a signal-to-noise that allow comparison to the spatial structures identified in our models. However, Fomalhaut's two companion stars are located much too far away, at a projected distance of 0.28 and 0.77 pc, respectively, to be considered for the scenario explored in the present study. 

HR\,4796\,A is not amongst this list of 7 resolved systems because the bound character of the nearby star HR4796B has not been established yet. It could, however, be a promising candidate, because its complex disc displays several similarities with our simulations. The first one is a narrow and bright main ring, located at $\sim75$\,au, which is surrounded by an extended halo of small particles that is dimmer than the one expected for an unperturbed disc around a single star \citep{lagr12} and could thus be the result of the perturbations of the M star HR 4796 B located at a projected separation of $\sim560$\,au. \cite{schn18} also identified a more distant bow-like structure, roughly located between HR 4796 A and HR 4796 B, which seems, at least qualitatively, reminiscent of the arc-like feature that we obtain in the direction of the companion when it emits a powerful stellar wind (Figs.~\ref{cutcm} and \ref{surfmasspr}). Interestingly, both the observed bow and our arc-like feature are followed by a sharp density drop in the direction of the companion. Another similarity is that the diffuse axisymmetric halo that starts at the main ring is, for both HR 4796 and our high-stellar-wind case, more extended in the anti-companion direction.
There are, however, some important differences. Firstly, the bow structure is located much closer to the primary star in the HR 4796 system, but closer to the companion star in our simulations. Secondly, the curvature of the bow is different: it is concave (when considered from the primary star) on the HR 4796 images but convex in our scenario. Last but not least, it is not clear whether the HR 4796 A \& B system is actually a physical binary. \cite{olof19} indeed computed the escape velocity of B with respect to A, using the parallaxes and proper motions provided by the second Gaia data release \citep{gaia18} as well as radial velocity measurements, and estimated that the two stars are probably not gravitationnally bound, even though additional data is required to settle the issue.

Conversely, we might look for resolved systems displaying structures similar to the ones obtained in our investigation, even if they aren't part of a known multiple system, because not all disc-bearing host stars have been thoroughly vetted for companion stars. In this respect, the disc having the most interesting geometry is clearly the one surrounding TWA7, with a narrow birth ring at $\sim25$\,au from which a single spiral arm is extending out to at least $\sim50\,$au \citep{olof18}. A potential explanation could be a yet undetected companion located in the prolongation of this arm, producing a pattern that is very similar to what we obtain in our nominal case (Fig.~\ref{e0im}). TWA7 is an M star with no radiation pressure able to push small grains outside the main debris ring, but a strong stellar wind (typical for young M stars) could have a similar effect. 
However, \cite{wahh13} did not find any bound companion in the star's vicinity, and mining the recent Gaia-DR2 catalogue \citep{gaia18} confirmed that all three stars that are angularly close to TWA7 are not located at the same distance, making the binary scenario less attractive.

\subsection{Signatures on the disc-integrated spectrum}

\subsubsection{colder SEDs}

An important and robust result is that, compared to a companion-free case, the system as a whole is depleted in small $s\lesssim 10\,\mu$m grains (Fig.~\ref{psde0}). This is mainly because these grains are the ones that are placed, by radiation pressure, on orbits reaching the unstable zone beyond the outer edge $r_{\rm{out}}$ of the PB ring. Even though the steady collisional production of these grains in the PB belt makes that a halo is still present beyond $r_{\rm{out}}$ (see Sec.\ref{trends}), this halo is more tenuous than in a companion-free case. 

This depletion of small grains affects the system's SED, but in a rather counter-intuitive way: the spectrum peaks at a larger wavelength than for the same PB ring with no companion (Fig.~\ref{sedc}), despite of the fact that there is actually \emph{less} matter in the outer $r>r_{\rm{out}}$ halo regions.
This is because small micron-sized grains have temperatures that strongly exceed the almost blackbody temperature of larger grains \citep[e.g.][]{theb19}, so that the contribution to the SED of these halo grains actually peaks at a shorter wavelength than that of the larger grains in the PB belt (Fig.~\ref{sede0})\footnote{An effect that is compounded by the fact that small grains become poor emitters at longer wavelengths}. As a consequence, their depletion, both in the PB ring and the halo, moves the SED towards larger wavelengths \footnote{The fact that, for the same PB-belt at 85au, the nominal SED appears to have a higher level than the no-companion SED in the $\geq100\mu$m region comes from the fact that the SEDs are renormalized to their maximum value. In absolute value, the two SEDs have comparable levels at these larger wavelengths }.
If we take as a reference parameter the wavelength $\lambda_{\rm max}$ at which the total SED peaks, then we find that, for our nominal case, binarity displaces $\lambda_{\rm max}$ from $\sim58\,\mu$m (value for a companion-free case) to $\sim78\,\mu$m. Considering the typical temperature $T_{\rm dust}=5100/\lambda_{\rm max}$ given by the Wien displacement law, this correspond to a temperature drop from $T_{\rm dust}\sim85$ K to $T_{\rm dust}\sim65$ K.

\subsubsection{explaining "unstable" debris discs}

\begin{figure}
\includegraphics[scale=0.5]{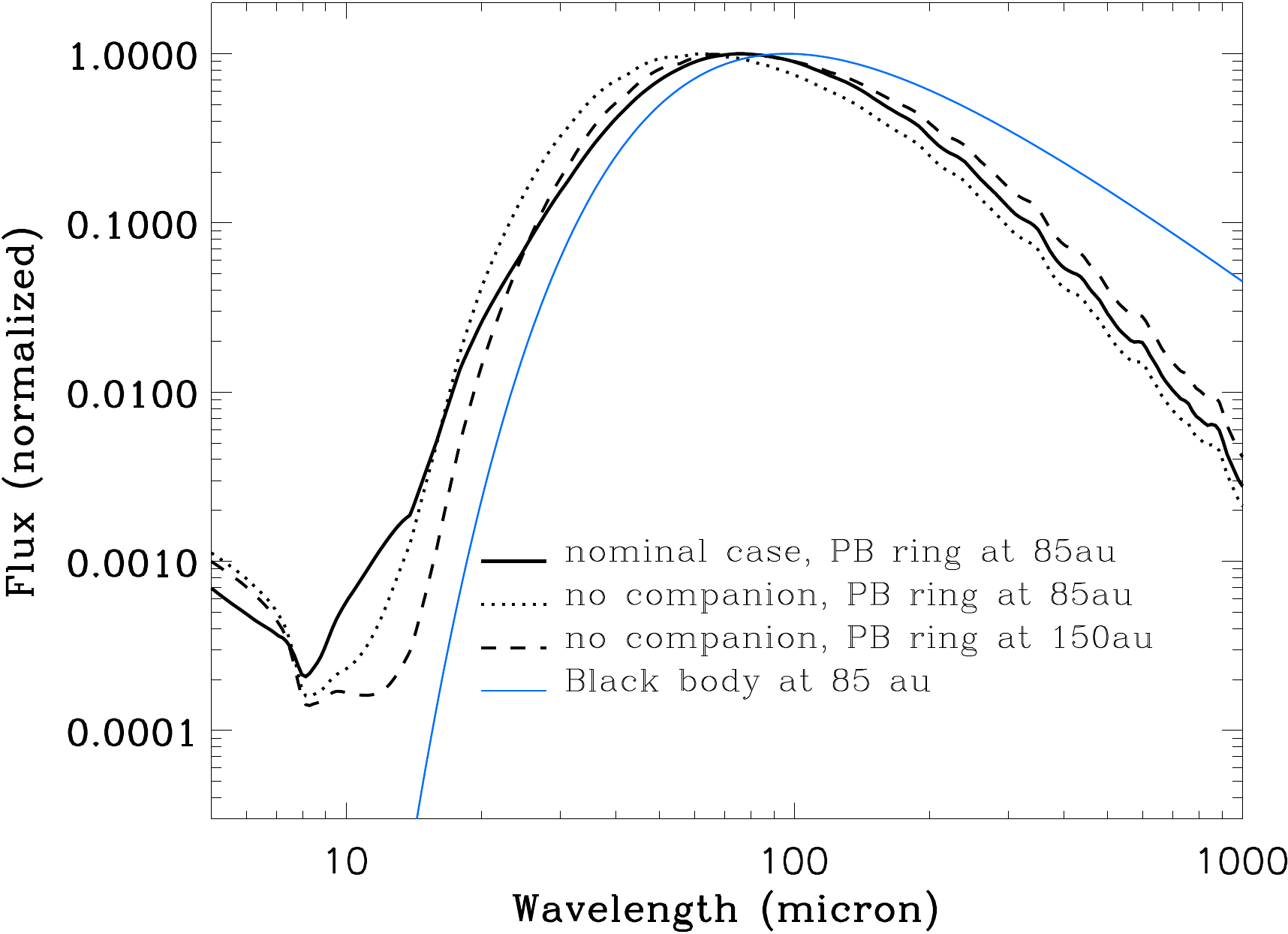}
\caption[]{Disc-integrated SED for the nominal case as well as for 2 companion-free cases: one where the PB ring is at the same distance, i.e., 85\,au for the centre of the ring, as in the nominal case, and one where the disc is moved outward by 75\%. We also plot, as a comparison, a blackbody profile at a distance of 85\,au from the central star.The different SEDs have all been normalized to their maximum value.}
\label{sedc}
\end{figure}

These results might help explaining the puzzling existence of a handful of discs-in-binaries whose SED-derived temperatures seemingly place the emitting dust in the dynamically unstable region between the two stars \citep{tril07,yelv19}. In TBO12 we suggested that this could be due to the small grains that do in fact populate these unstable $r>r_{\rm{out}}$ regions, but without quantifying their impact on the SED. 
While we identify this halo of unstable grains in the present study, and show that it even could be visible on resolved images (see Sec.\ref{resolved}), we find that it is not the presence of these halo grains but, on the contrary, their \emph{under-abundance} (with respect to a companion-free case) that is the key result here: less of these hot grains actually displaces the SED towards longer wavelengths than for the same PB belt without a stellar companion. In other words, a disc in a binary might mimic the thermal behaviour of a more distant disc around a single star. In fact, because of this depletion of small grains, the departure from a blackbody case is much more limited than for a disc around a single star (Fig.~\ref{sedc}).

This results supports the conclusion of \cite{yelv19} that, while nine out of their 38 discs appear to be in the unstable region based on simple estimate of the dust radial location $r_{\rm dust}$ (using the empirical relations by \citealp{pawe14} for a \emph{single star} environment), the proportion of genuinely unstable discs is probably much lower, notably because of large uncertainties regarding $r_{\rm dust}$. Our results show that using disc radii estimates derived for single stars 
can indeed overestimate the value of over $r_{\rm dust}$ by as much as $\sim75\%$. 
We note that, if we consider all unstable discs (discarding the very different cases that are circumbinary discs) displayed in Fig.~5b of \cite{yelv19} and reduce their $r_{\rm dust}$ (called "$r_{\rm disc}$" in that paper) by the $\sim$75\% difference we find between $r_{\rm dust}$ and $r^{\rm single}_{\rm dust}$, most of them either move out of the unstable region or end up at the edge of it. 

Our results could also explain why, for the 2 "unstable" discs for which resolved images are available (HD 181296 and HD 207129), there is a discrepancy between $r_{\rm dust}$ derived from the SED and the observed location of the disc, the latter being always closer to the central star than the former.
We chose, however, to remain cautious when assessing specific cases, as there are, most of the time, several crucial parameters that are poorly constrained, notably the binary's orbit, which is very often unknown (only the projected distance between the two stars being constrained) and leaves large uncertainties regarding the perturbing effect of the companion star.

\subsection{Detectability of the secondary ring}\label{secring}

Let us investigate whether the secondary ring which can clearly be seen on the normalised luminosity maps shown in the paper, would be detectable with current instruments. First, we will look into the possibility of resolving the ring and then look into whether an excess compared to the photosphere emission can be detected when the ring cannot be resolved.

The secondary ring observed in the simulations is typically located between $15$ and $30$\,au from the companion star. For the most favourable case $\mu=1$ (see Fig.~\ref{surfmass}b), we estimate that the ring has an IR fractional luminosity of $5\times 10^{-5}$. Its disc-integrated flux in scattered light would peak at $\sim 5$ mJy at $\lambda\sim0.5\,\mu$m if it were to be observed at a $20$\,pc distance (the approximate distance of the archetypal $\beta$ Pic system), and it would have a typical on-sky radius between $0.75$ and $1.5$\arcsec.

To check whether SPHERE could detect such a ring, we generated an image, in total intensity, at $1.6$\,$\mu$m with a pixel size corresponding to that of the SPHERE/IRDIS instrument ($0.01226\arcsec$), and we find that, for the $\mu=1$ case seen at a typical distance of $20$\,pc, the mean flux density per unit surface in the secondary ring is $350$\,$\mu$Jy/arcsec$^2$, with a peak value reaching $1.8$\,mJy/arcsec$^2$. Unless the secondary disk is highly inclined, the best strategy to detect such a disk would be differential polarimetric imaging (DPI), as any disks with an inclination smaller than $\sim60^{\circ}$ would not be detectable using the angular differential imaging technique (see \citealp{mill12}). During DPI observations, the linear polarization is measured along two sets of orthogonal directions to measure the $Q$ and $U$ components of  the Stokes vector. Given that the stellar light is unpolarized, its contribution is expected to be identical in the $Q$ and $U$ images, and can therefore be efficiently subtracted. Even though the noise in the inner $0.15$\arcsec of the final images can still be significant, at distances larger than $\sim 0.5$\arcsec, the contribution of the stellar point-spread function is negligible. Examples ranging from very bright to very faint disks would be; HR\,4796\,A \citep[][$1.07$\arcsec]{mill19}, $49$\,Cet \citep[][$\sim 2.2$\arcsec]{choq17}, and TWA\,7 \citep[][$0.73$\arcsec]{olof18}. Making reliable predictions on whether a disk can be detected in DPI is very challenging as it depends on the extent of the disk, its inclination, radius, or the degree of polarization of the small dust grains (i.e., the dust properties such as the minimum grain size and porosity). Nonetheless, based on previous detection and the typical surface brightness of those disks, we can estimate whether the secondary disk could be detected or not if it were at a distance of $20$\,pc. For instance, \citet{choq17} detected the extended and faint disk around $49$\,Cet with SPHERE, and reported a surface brightness of $\sim 150$\,$\mu$Jy/arcsec$^2$ (in total intensity). The face-on disk around TWA\,7 was detected with SPHERE at the level of $\sim 100$\,$\mu$Jy/arcsec$^2$ in DPI mode (revised reduction of the dataset presented in \citealp{olof18}, Ren et al., in prep, private communication). As demonstrated in \cite{espo20}, debris disks with surface brightness of the order of hundreds of $\mu$Jy/arcsec$^2$ can be detected with second generation instruments such as the \textit{Gemini Planet Imager} (GPI) in DPI \citep{perr15}. This suggests, that a detection of the disk around the secondary star with SPHERE or GPI would be possible, though not trivial \footnote{Note that, because particles passing too close to the secondary star have been removed in our simulations, the luminosity of this secondary disc is likely to be underestimated, so that our detection criteria is probably conservative}.
Thanks to the exquisite resolution of SPHERE (Strehl ratio as high as $90$\% and a point spread function with a full width at half maximum of a few pixels, e.g. $0.046$\arcsec, \citealp{mill19}), if detected, a ring with a typical radius between $0.75\arcsec$ and $1.5\arcsec$ would be well resolved at 20pc, across tens of resolution elements. Furthermore, the two stars would be separated by roughly $19\arcsec$ (for $\mu=1$), meaning that, given that the IRDIS instrument field of view is of 11\arcsec \citep{dohl08}, the secondary ring could be observed with minimal light pollution of the primary star. In the close-future, the JWST and the ELT would probably be able to detect and image this type of faint and compact secondary rings around nearby companion stars given that the primary star can be shifted outside of the field of view or hidden behind a coronagraph. The aforementioned seven systems with resolved \emph{primary} discs would be an obvious target list for attempting such challenging detections.

\begin{figure}
\includegraphics[scale=0.5]{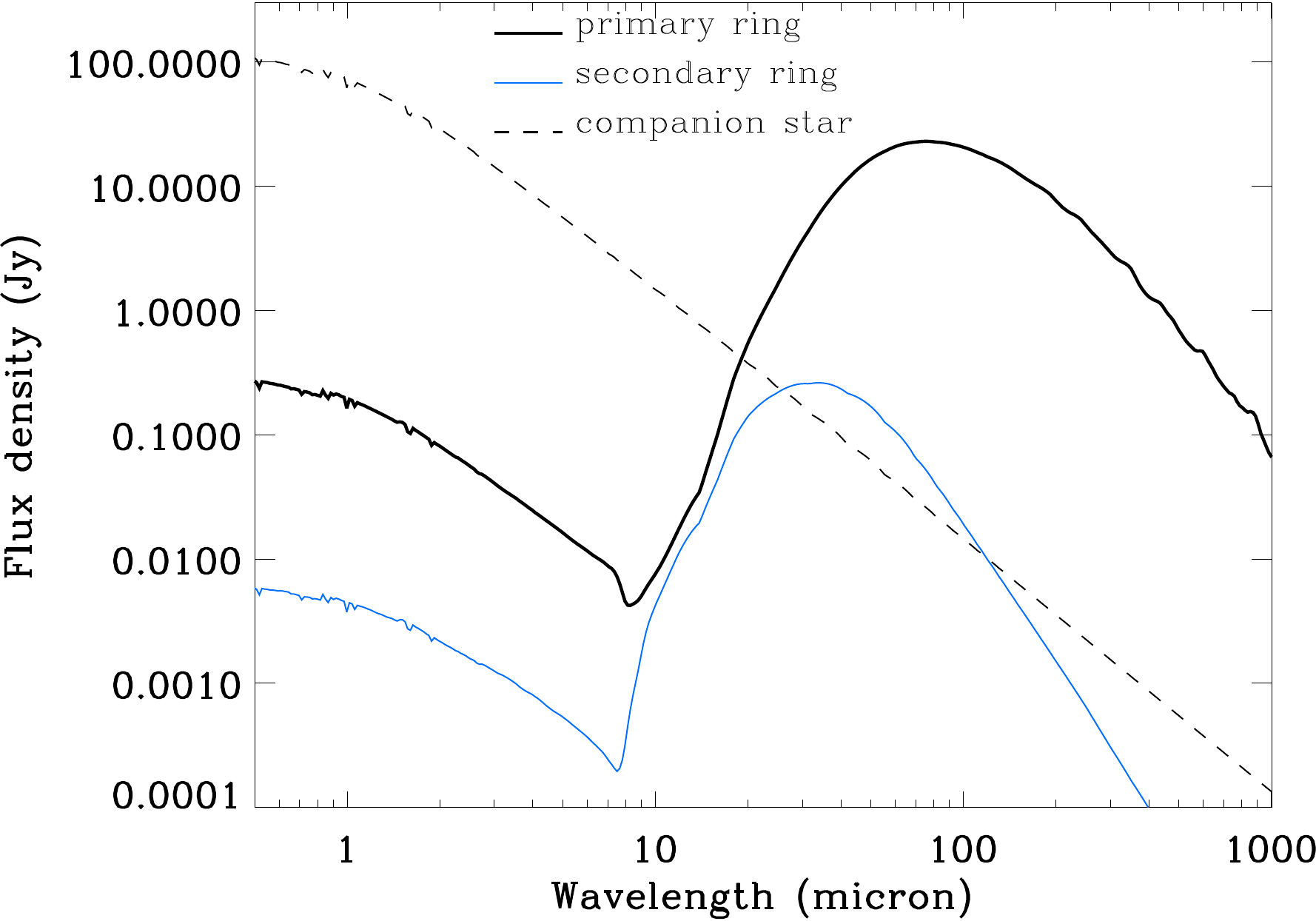}
\caption[]{Massive companion case ($\mu=1$): absolute SED for a system viewed at a distance of 20\,pc}
\label{fluxh}
\end{figure}

In the mid-to-far IR, the secondary ring would probably be very difficult to resolve with current or past instruments (e.g. Herschel/Spitzer). However, even if the resolution is not enough to image the ring, it might be sufficient to separate its contribution from that of the primary ring. This would require a resolution of $\sim10\arcsec$ for a system at $20$\,pc, and in this case the secondary ring's IR excess might be photometrically detected on the SED. This can be seen for the most favourable case ($\mu=1$) displayed in Fig.~\ref{fluxh}, where the SED of the secondary ring peaks at 0.4 Jy at around 30 microns and lies above that of the companion star up to a bit more than 100 microns in wavelength. This value of flux density could have been readily detected by Herschel \citep{eiro13}, Wise or Spitzer. This disc could also be detected by SOFIA/FORCAST in a short integration time. This could also be done at a range of wavelengths with e.g. Origins Space Telescope in the future from the mid-to-far infrared.

Future observations might thus reveal a complex situation, where binaries showing both circumprimary and circumsecondary discs could either correspond to two separate "native" discs or to only one genuine collisional disc that is feeding the other. To tell these two scenarios apart one might have to look at the relative luminosity and extent of the 2 discs, with a non-native secondary disc always being more compact and fainter than its primary counterpart. Careful analysis of the secondary's SED, and especially its slope at larger wavelengths, could also provide crucial information: a non-native secondary disc is indeed not linked to a local parent body belt and is strongly depleted in grains beyond the $\sim10\,\mu$m size.

\section{Summary and Conclusion}\label{ccl}

Using the state-of-the-art DyCoSS code, we investigate the maximum possible effect a companion star can have on a debris disc, i.e., when the outer edge of its dust-producing parent body (PB) belt coincides with the instability limit $r_{\rm{crit}}$ due to companion perturbations (meaning that the companion star is as close as possible to the belt without destroying it). Our main results can be summarized as follows:

\begin{itemize}
\item We confirm preliminary results obtained in TBO12, which is that the combined effect of collisional production in the PB belt, stellar radiation pressure (or stellar wind) and dynamical perturbations is able to sustain a significant level of dust in the unstable region beyond $r_{\rm{crit}}$. This unstable halo is made of micron-sized grains and should be detectable on resolved images in the visible up to the mid-IR. 

\item Companion perturbations also create a halo interior to the PB ring. This inner halo dominates the system's total flux in the $\lambda\sim10-20\,\mu$m range, but remains difficult to observe because this is the wavelength domain where the system is the dimmest.

\item Depending on the binary configuration, several asymmetric features are observed, the most prominent one being a spiral arm extending from the PB ring all the way to the companion. The arm precesses with the companion for circular binaries, but is only present close to periastron passages for high-$e$ binaries.

\item For most explored cases, we obtain a compact secondary ring around the companion star, which feeds off the unstable circumprimary halo. As such, this secondary ring is non-native and is not associated to the collisional grinding of a local belt of larger progenitors. It is mostly made of grains in the $\sim1-10\,\mu$m size range. For bright circumprimary discs around A stars located at distances of a few tens of pc, this could be detected and resolved with instruments like SPHERE. This secondary disc could also be detectable in photometry provided that its contribution can be separated from that of the primary disc. 

\item Among the 7 unambiguously resolved known discs in binaries only Fomalhaut presents images that have high-enough resolution and signal-to-noise to allow comparisons to our models. However, its companion stars are much too far away from the main disc to correspond to the scenario that we explored. HR 4796 could be an interesting candidate, with structures partially matching those identified here and a potential companion star at a compatible distance from its main disc, but the bound character of the HR 4796 A and B pair is far from being established. The debris disc displaying most similarities with our predictions is TWA7, but so far no companion star has been found in its vicinity with very constraining upper limits.

\item When compared to a companion-free case, the system as a whole is significantly depleted in small, $s\lesssim10\,\mu$m grains, whose orbits almost always enter the dynamically unstable $r>r_{\rm{crit}}$ region.
This depletion of small grains, which are hotter (even in the outer halo) than the larger grains in the PB belt, shifts the system-integrated SED towards longer wavelengths. As such, the disc as a whole appears colder than for the same PB belt around a single star.

\item This shift towards longer wavelengths could explain why, when using empirical rules derived for discs around single stars, the SED-derived location $r_{\rm disc}$ of some unresolved discs in binaries appear to fall in the unstable region between the two stars. We suggest that this apparent paradox could be due to an overestimate of $r_{\rm disc}$ that could be as high as $\sim75$\% when not accounting for binarity-induced effects.

\end{itemize}

\begin{acknowledgements}
The authors thank the reviewer for very constructive remarks that helped improve the paper. J.\,O. acknowledges financial support from the ICM (Iniciativa Cient\'ifica Milenio) via the N\'ucleo Milenio de Formaci\'on Planetaria grant, from the Universidad de Valpara\'iso, and from Fondecyt (grant 1180395).
\end{acknowledgements}

{}

\clearpage

\end{document}